\documentclass[twocolumn,english,aps,pra,reprint,noeprint,superscriptaddress,nofootinbib]{revtex4-2}
\usepackage{physics}
\newcommand{\FAC}{\texttt{FAC}}

\newcommand{\HULLAC}{\texttt{HULLAC}}
\usepackage{amsmath}
\usepackage{graphicx}
\usepackage{multirow}
\usepackage{bm}
\usepackage{tikz}
\usepackage{xcolor} 
\usepackage{hyperref}
\usepackage[capitalize]{cleveref}

\usepackage{amsmath}
\usepackage{amssymb}
\usepackage{hyperref}
\usepackage[defaultcolor=teal, draft]{changes}
\usepackage{xcolor}
\hypersetup{
	colorlinks,
	linkcolor={blue!70!black},
	citecolor={blue!70!black},
	urlcolor={blue!70!black}
}
\usepackage{orcidlink}
\newcommand\ion[2]{\text{#1\,\textsc{\lowercase{#2}}}}	

\usetikzlibrary{shapes.geometric, arrows}
\tikzstyle{startstop} = [rectangle, rounded corners, minimum width=3cm, minimum height=1cm, text centered, draw=black, fill=red!30]
\tikzstyle{process} = [rectangle, minimum width=3cm, minimum height=1cm, text centered, draw=black, fill=orange!30]
\tikzstyle{decision} = [diamond, minimum width=3cm, minimum height=1cm, text centered, draw=black, fill=green!30]
\tikzstyle{arrow} = [thick,->,>=stealth]

\begin{document}


\title{Systematic Bayesian Optimization for Atomic Structure Calculations of Heavy Elements}

\author{Ricardo Ferreira da Silva\,\orcidlink{0000-0003-3030-0496}}\email{rfsilva@lip.pt}

\affiliation{Laboratório de Instrumentação e Física Experimental de Partículas (LIP) \\ Av. Prof. Gama Pinto 2, 1649-003 Lisboa, Portugal}
\affiliation{Faculdade de Ciências da Universidade de Lisboa \\ Rua Ernesto de Vasconcelos, Edifício C8, 1749-016, Lisboa, Portugal}

\author{Andreas Fl{\"o}rs\,\orcidlink{0000-0003-2024-2819}}
\affiliation{GSI Helmholtzzentrum f{\"u}r Schwerionenforschung \\
Planckstra{\ss}e 1, D-64291, Darmstadt, Germany}

\author{Luís Leitão\,\orcidlink{0009-0002-8826-9284}}
\affiliation{Laboratório de Instrumentação e Física Experimental de Partículas (LIP) \\ Av. Prof. Gama Pinto 2, 1649-003 Lisboa, Portugal}
\affiliation{Faculdade de Ciências da Universidade de Lisboa \\ Rua Ernesto de Vasconcelos, Edifício C8, 1749-016, Lisboa, Portugal}

\author{José~P.~Marques\,\orcidlink{0000-0002-3797-3880}}
\affiliation{Laboratório de Instrumentação e Física Experimental de Partículas (LIP) \\ Av. Prof. Gama Pinto 2, 1649-003 Lisboa, Portugal}
\affiliation{Faculdade de Ciências da Universidade de Lisboa \\ Rua Ernesto de Vasconcelos, Edifício C8, 1749-016, Lisboa, Portugal}

\author{Gabriel~Mart{\'i}nez-Pinedo\,\orcidlink{0000-0002-3825-0131}}
\affiliation{GSI Helmholtzzentrum f{\"u}r Schwerionenforschung \\
Planckstra{\ss}e 1, D-64291, Darmstadt, Germany}
\affiliation{Institut f{\"u}r Kernphysik (Theoriezentrum), Technische Universit{\"a}t Darmstadt \\
Schlossgartenstra{\ss}e 2, D-64289, Darmstadt, Germany}

\author{Jorge~M.~Sampaio\, \orcidlink{0000-0003-4359-493X
}}
\affiliation{Laboratório de Instrumentação e Física Experimental de Partículas (LIP) \\ Av. Prof. Gama Pinto 2, 1649-003 Lisboa, Portugal}
\affiliation{Faculdade de Ciências da Universidade de Lisboa \\ Rua Ernesto de Vasconcelos, Edifício C8, 1749-016, Lisboa, Portugal}

\begin{abstract}
This study presents a novel optimisation technique for atomic structure calculations using the Flexible Atomic Code, focussing on complex multielectron systems relevant to $r$-process nucleosynthesis and kilonova modelling. We introduce a method to optimise the fictitious mean configuration used in the Flexible Atomic Code, significantly improving the accuracy of calculated energy levels and transition properties for lanthanide and actinide ions.
Our approach employs a Sequential Model-Based Optimisation algorithm to refine the fictitious mean configuration, iteratively minimising the discrepancy between calculated and experimentally determined energy levels. We demonstrate the efficacy of this method through detailed analyses of Au II, Pt II, Pr II, Pr III, Er II, and Er~III, representing a broad range of atomic configurations.
The results show substantial improvements in the accuracy of the calculated energy levels, with average relative differences to the NIST data reduced from 20-60\% to 10\% or less for the ions studied. Transition wavelength calculations exhibit good agreement with experimental data, with about 90\% of the calculated values falling within 10\% of measurements for Pr and Er ions. While improvements in transition probability calculations are observed, the calculated radiative transition probabilities (log($gf$) values) still show significant discrepancies compared to the experimental data, with root mean square deviations of approximately 1.1-1.4 dex for Pr and Er ions.
We extend our optimisation technique to systematic calculations of singly and doubly ionised lanthanides, achieving accuracies comparable to or surpassing those of \textit{ab-initio} atomic structure codes. The method's broad applicability across the lanthanide series demonstrates its potential for enhancing opacity calculations and spectral modelling in astrophysical contexts.
\end{abstract}

\maketitle

\section{Introduction}

Neutron star mergers have been confirmed to be sites of rapid neutron capture ($r$-process) nucleosynthesis, producing heavy elements beyond iron \cite{Lattimer1974y, Lattimer1976c, Eichler1989l, Rosswog1999m, Freiburghaus1999x}. The electromagnetic counterpart to these events, known as kilonovae, provides a unique opportunity to study the creation and distribution of $r$-process elements~\cite{Metzger.Martinez-Pinedo.ea:2010}. 
Kilonova AT2017gfo has been extensively studied~\cite{Abbott2017y, Cowperthwaite2017z, Kasen2017z, Smartt2017t, Villar2017k}, indicating the presence of a wide range of $r$-process elements \cite{Tanaka2018f, Wollaeger2018x, Bulla2018y, Watson2019e, Kawaguchi2021q, Domoto2021v, Gillanders2021c, Domoto2022m, Gillanders2022n, Hotokezaka2023v, Sneppen2023h, Tarumi2023i}.

The modelling and interpretation of kilonova observations critically depend on accurate atomic data for $r$-process elements. Radiative transfer models require a comprehensive understanding of level energies and bound-bound atomic transitions, which account for most of the photon opacity in $r$-process-enriched ejecta \cite{Kasen2013w}. Beyond one day after the merger, the ejecta mainly comprise neutral to four-times ionised atoms \cite{Shingles2023s}. Lanthanide and actinide ions, characterised by their high-level density due to complex atomic structures with open valence $f$-shells \cite{Fritzsche2022p}, are expected to contribute significantly to the opacity of the ejecta through numerous transitions. However, experimental data are available for only a small subset of these elements \cite{Kramida2021w}. This scarcity of experimental data, coupled with the inherent complexity of these elements, poses significant challenges in atomic structure calculations. Consequently, theoretical atomic structure calculations are essential to provide a complete atomic data set of opacities for all ions of modelling interest, including particularly challenging actinides. Previous work has made significant progress in the calculation of atomic data for lanthanides and some actinides
\cite{Tanaka2020t, Fontes2020g, Radziute2020n, Radziute2021s, Silva2022i, Fontes2022l,Pognan2022o, Flors2023a}, demonstrating the critical role of accurate atomic data in understanding kilonova spectra and light curves. Nevertheless, highly accurate atomic structure calculations for these complex systems remain a challenging and time-consuming task, underscoring the need for continued refinement of computational methods.

However, recent observational advances have provided crucial insight into the composition of kilonovae. Multiple independent analysis of the spectra have identified strontium features (Sr, $Z = 39$) in the first days after the merger \cite{Watson2019e, Domoto2021v}. Additional detections of elements on the left side of the periodic table, including zirconium (Zr, $Z = 40$) and lanthanides lanthanum (La, $Z = 57$) and cerium (Ce, $Z = 58$), have also been reported \cite{Domoto2022m}. Preliminary efforts to identify gold and platinum have not succeeded, underscoring the persistent difficulties in identifying some of the most massive $r$-process elements within kilonova spectra \cite{Gillanders2021c,Mccann.Mulholland.ea:2024,Pognan.Wu.ea:2024}.

Many approaches exist to perform atomic structure calculations, each offering different balances between accuracy and computational efficiency. The multiconfiguration Dirac-Fock method, implemented in codes such as \texttt{GRASP2018}  \cite{FroeseFischer2019d} and \texttt{MCDFGME} \cite{OGorceix1987g}, can achieve high accuracy by optimising selected wavefunctions or minimising the energy function for individual or groups of levels. However, these calculations often come at the cost of significant computational time. Alternatively, codes like \texttt{AUTOSTRUCTURE} \cite{Badnell2016c} and the Cowan suite of codes \cite{Cowan1981n}  employ very large configuration interaction (CI) expansions to achieve similar accuracy. These methods rely on radial wavefunctions derived from model potentials with scaling parameters, which are then fine-tuned based on available experimental data. Although this approach can help with accuracy, it requires careful adjustment and validation against known spectroscopic data. Recent work has shown promise in applying Bayesian machine learning techniques to optimise the scaling parameters of Slater-type orbitals used in the \texttt{AUTOSTRUCTURE} code. This technique was applied in the optimisation of orbital radial wavefunctions for neutral beryllium calculations \cite{Mendez2020i,Mendez2021g}.

Finally, other codes such as \HULLAC\, \cite{Bar-Shalom2001d} and the Flexible Atomic Code (\FAC) \cite{Gu2008f}, rely on a local spherically averaged central potential to describe electron-electron and electron-nucleus interactions. Typically, this potential is derived from a fictitious mean configuration (FMC), assuming a uniform electron distribution. Although this approach simplifies calculations, it can introduce significant errors, particularly for near-neutral heavy elements with complex electron configurations, such as those found in the lanthanide and actinide series owing to the presence of open $d$ and $f$ shells.

The purpose of this study is to enhance the accuracy of atomic structure calculations by optimising the FMC. By refining the occupancy numbers of valence electrons, we seek to improve the agreement between the calculated energy levels and experimentally determined reference values. This optimisation employs a Sequential Model-Based Optimisation (SMBO) algorithm, implemented using the \texttt{scikit-optimize} (\texttt{skopt}) library in Python \cite{Head2021z}. The SMBO algorithm is particularly effective for high-dimensional optimisation problems, balancing the exploration of new parameter spaces with the exploitation of known low-error regions.

Our methodology takes advantage of a carefully selected set of low-lying energy levels, either from experimental data or accurate \textit{ab initio} calculations as benchmarks. These benchmarks guide the iterative refinement of the FMC, ultimately producing an optimised potential that significantly improves the fidelity of atomic structure calculations. The optimised FMC approach is not only applicable to lanthanides and actinides but can be generalised to other elements and ionisation states, providing a robust tool for high-precision atomic data generation.

To demonstrate the effectiveness and versatility of our method, we present detailed analyses for selected ions across different elements and ionisation states. Specifically, we focus on \ion{Au}{II}, \ion{Pt}{II}, \ion{Pr}{II}, \ion{Pr}{III}, \ion{Er}{II}, and \ion{Er}{III} as case studies. These ions were chosen to represent a range of atomic structures and complexities. \ion{Au}{II} and \ion{Pt}{II} serve as benchmarks for our optimisations near the third $r$-process peak, which are of particular interest in kilonova spectra \cite{Gillanders2021c}. \ion{Pr}{II}, \ion{Pr}{III}, \ion{Er}{II}, and \ion{Er}{iii} demonstrate the performance of the method across the lanthanide series, showcasing its applicability to different ionisation states and varying $4f$ shell complexities. By examining these diverse cases, we aim to illustrate the broad applicability of our optimised FMC approach and its potential to significantly enhance the accuracy of atomic structure calculations across the periodic table.

Our primary objective is to provide comprehensive atomic data for all lanthanide ions, addressing a critical need in the field of astrophysics and atomic physics. This data is essential for accurate modelling of kilonova spectra and for understanding the nucleosynthesis of heavy elements in neutron star mergers. By improving both the accuracy and efficiency of these calculations, we aim to enable more precise astrophysical modelling and potentially facilitate the identification of additional elements in future kilonova observations. Our optimisation procedure not only improves the accuracy of atomic structure calculations but also provides a crucial foundation for reliable level identification, which is essential for any subsequent energy level calibration against experimental data.

In the following sections, we will detail the computational procedures, describe the optimisation methodology, present the results of our calculations for the selected ions, and discuss their implications for astrophysical modelling and other applications requiring precise atomic data. We will also explore how this improved atomic data can contribute to our understanding of $r$-process nucleosynthesis and potentially aid in the identification of heavy elements in future kilonova observations.

\section{Atomic Structure Calculations}

The calculations in this work were performed using \FAC\ (Flexible Atomic Code) \cite{Gu2008f}. \FAC\ is an integrated software package for the calculation of various atomic structure and collisional processes based on a relativistic configuration interaction (RCI) method. In this method, the eigenstates are represented by atomic state functions $\Psi$ which are constructed from a superposition of $i=1,\cdots , N_{\texttt{CSF}}$ configuration state functions (CSFs) $\varphi_i$ with the same total angular momentum $J$ and parity $P$,
\begin{equation}
\Psi (\gamma  J M_J P)=\sum_i^{N_{\texttt{CSF}}}c_i \; \varphi_i ( \gamma_i  J M_J P), 
\label{eq:wavefunc}
\end{equation}
where the $\gamma_i$ encompasses all the remaining relevant information to define each CSF uniquely.

The mixing coefficients $\{c_i\}$ are obtained by solving the eigenvalue problem $\mathbf{H}\mathbf{c} = E\mathbf{c}$, with $\mathbf{c} = (c_1, c_2, \ldots , c_{N_{\texttt{CSF}}} )^t$. The eigenvalues obtained from the diagonalisation of the Hamiltonian matrix $\mathbf{H}$ are therefore the best approximation for the energies in the space described by the selected CSFs. Although increasing the number of CSFs would improve the wave functions and the expected accuracy of the atomic energy levels (AELs), this improvement must be balanced against computational cost. Therefore, we determine an optimal set of CSFs by examining the energy level convergence with an increasing configuration basis.

\FAC\ employs a variant of the conventional Dirac-Fock-Slater method to compute one-electron radial functions. Radial components are determined by solving the coupled Dirac equations self-consistently for a local central potential $V(r)$, following standard relativistic formulations \cite{Cowan1981n}. A unique potential is used to derive the radial orbitals for the construction of basis states. As such, orthogonality is automatically ensured.

The local central potential includes contributions from both nuclear and electron-electron interactions. The nuclear contribution is modelled using a uniformly charged sphere with a radius determined by the atomic mass \cite{Gu2008f, CHERNYSHEVA1999232}. For the electron-electron contribution, in the current version of \FAC, contrary to earlier documentation, a Dirac-Fock-Slater (DFS) potential is implemented \cite{Badnell2024k}
\begin{equation}
V_{ee}(r) = \sum_{\beta} \omega_{\beta} Y^0_{\beta,\beta}(r) - A \rho(r)^{1/3}
\label{eq:dfs_potential}
\end{equation}
where $\beta={n \kappa}$ defines the principal quantum number $n$ and relativistic quantum number $\kappa$ for each subshell and  
\begin{equation}
    Y_{\beta \eta}^{\lambda}(r) = \int \frac{r^{\lambda}_{<}}{r^{\lambda+1}_{>}} \rho_{\beta}(r^\prime) \, d r^\prime
\end{equation}
with $r^{\lambda}_{<} = \min(r,r^\prime) $, $r^{\lambda+1}_{>}= \max(r, r^\prime)$.

The first term represents the direct Coulomb interaction, spherically averaged over the bound electron states. Here $\omega_\beta=\omega_{n \kappa}$ are the occupation numbers for each subshell subject to $\sum \omega_{\beta} = N_e$, where $N_e$ is the total number of electrons. An intrinsic feature of this Dirac-Slater approach is the inclusion of electron self-interaction, which manifests itself in the asymptotic behaviour $rV(r) \to N_e$ as $r \to \infty$, rather than the physically expected $N_e-1$. While the self-interaction in the direct term is partially compensated for by including self-exchange in the second term of \Cref{eq:dfs_potential}, \FAC\, additionally implements the Latter cut-off \cite{PhysRev.99.510}
\begin{equation}
V(r) = \min\left(V(r), \frac{N_e-1}{r}\right)
\label{eq:latter}
\end{equation}
to ensure the correct asymptotic behaviour. The local exchange potential is given in terms of the total spherically averaged electron number density
\begin{equation}
\rho(r) = \frac{1}{4\pi r^2} \sum_{\beta} \omega_{\beta} \rho_{\beta}(r).
\label{eq:density}
\end{equation}
The coefficient $A$ is simply a numerical factor pre-optimized for the ground configuration of each ion. This same approach has also been adopted in previous studies by Sampson \textit{ et al.}. (2009) \cite{Sampson2009s}.

These occupation numbers $\omega_{\beta}$ are derived from a single fictitious mean configuration (FMC), ensuring a unique potential $V(r)$ for all electrons. Various methods can be used to determine the FMC occupation numbers. The \FAC\ manual recommends distributing the occupation number of the electrons in the valence shells of the corresponding valence complex. Another approach involves splitting the occupancy number among a specific set of considered transitions, although this method is less computationally efficient \cite{Sampson1989a, Zhang1989q}. Both techniques have demonstrated effectiveness for highly ionised ions, where the wide level spacing and average screening effects yield reasonably accurate results. 

Although these techniques are effective for highly ionised systems where the level spacing is wide and average screening effects dominate, they become less reliable for near-neutral ions, particularly lanthanides and actinides with open $4f$ and $5f$ shells. The exceptionally high density of energy levels in these systems, exceeding even the $d$ block elements, makes the potential particularly sensitive to the screening contributions from electrons across different principal quantum numbers. This sensitivity motivates the optimisation procedure described in the following section.

\section{FMC optimisation procedure}

In this section, we present a method for optimising the local central potential in \FAC, calculations by automatically adjusting the occupancy numbers of valence electrons. The method takes advantage of the relationship between the electron occupancy numbers $\omega_{\beta}$ and the local central potential $V(r)$, as described in \cref{eq:dfs_potential}. This relationship, while central to the accuracy of the calculations, cannot be directly determined. Instead, we treat it as a black-box function $\mathcal{F}: \{\omega_{\beta}\} \rightarrow V(r)$, which we seek to optimise by minimising the discrepancy between the calculated energy levels and the experimental reference data.

For near-neutral lanthanide and actinide ions, we focus on optimising the occupancy numbers of four key valence shells: the $4f$, $5d$, $6s$ and $6p$ orbitals for lanthanides, and $5f$, $6d$, $7s$ and $7p$ orbitals for actinides. These shells are chosen because they are close in energy and exhibit significant configuration mixing, leading to important contributions to the electron density in the valence region. The $f$ and $d$ shells are particularly relevant because of their role in most of the low-lying configurations that dominate the atomic spectra.

The optimization is implemented using the Sequential Model-Based Optimisation (SMBO) algorithm through the \texttt{scikit-optimize} library in Python~\cite{Head2021z}. This approach is particularly effective for our four-dimensional optimization problem, as it efficiently balances the exploration of new parameter spaces with the exploitation of known low-error regions. To maintain physical relevance of our solutions, we constrain the occupancy numbers within specific bounds: the $f$-shell is restricted between $n_e - 2$ and $n_e + 2$, where $n_e$ is the number of electrons in the ground configuration's $f$ shell, allowing for physically reasonable excitations, while the outer $p$ shells are limited between 0 and 1 reflecting their typical occupancy in excited configurations.

To evaluate different potentials produced by $\mathcal{F}$, we employ a sequential model-based optimisation (SMBO) approach. At each iteration, $\mathcal{F}$ generates a potential $V(r)$ from a given set of $\{\omega_{\beta}\}$. This potential is used in the \FAC\ calculations to produce energy levels, which are then compared to the experimental data. We explored various statistical measures for this comparison, including mean absolute deviation, weighted average deviation, and level density. However, the limited and nonuniform availability of experimental uncertainties for lanthanide and actinide energy levels made some approaches impractical. After testing, we found that a weighted root mean square deviation (WRMSD) provided the most reliable convergence
\begin{equation}
   \mathcal{L}_{\text{weighted}} = \sqrt{\frac{1}{N} \sum_{i=1}^N e^{-E^{\text{ref}}_i/kT} (\Delta E_i)^2},
\end{equation}
where $\Delta E_i$ is the difference between the calculated and reference energy levels. The Boltzmann factor $e^{-E^{\text{ref}}_i/kT}$ with $T=5000$ K accounts for the thermal population distribution under LTE conditions, giving higher weight to lower-lying levels that contribute most significantly to opacity in kilonova. We note that as long as the temperature is representative of an LTE regime in kilonovae, small variations in the specific value have minimal impact on the optimization results. Using A-values or oscillator strengths as evaluation metrics was considered but rejected as it would require computing these values at each iteration, significantly increasing the computational time.

It is important to note that perfect agreement for all energy levels simultaneously cannot be guaranteed, as this is fundamentally limited by the local central potential approximation used in \FAC. Different energy levels might require different optimisation parameters for the best agreement, creating an inherent trade-off in the optimisation process. This limitation of the model is particularly evident in cases with strong configuration mixing, where the mean-field approach might not capture all correlation effects adequately. Our optimisation strategy therefore aims to find the best compromise solution that improves overall agreement while maintaining physical relevance.

The optimisation process requires balancing exploration of unknown regions with exploitation of known good solutions. This balance is achieved through three complementary acquisition functions: Lower Confidence Bounds (LCB), Expected Improvement (EI), and Probability of Improvement (PI). These functions are computed concurrently, with LCB focussing on the trade-off between predicted mean and uncertainty, EI targeting potential improvements over the current best solution, and PI evaluating the probability of finding better solutions. A GP-Hedge portfolio strategy \cite{10.5555/3020548.3020587} probabilistically selects among them using a softmax function, ensuring robust exploration of the parameter space while efficiently converging to optimal solutions.

Optimisation begins with the initial values $\{\omega_{\beta}\}$ derived from the first few energetically ordered configurations, ensuring convergence to physically meaningful values. Subsequent points are chosen using Latin hypercube sampling, which ensures efficient coverage of the parameter space by stratifying the sampling regions. A Gaussian surrogate model approximates $\mathcal{F}$, creating an effective four-dimensional surface that guides optimisation. The model is updated after each evaluation to incorporate new information and guide the selection of subsequent points in the parameter space. The process typically converges within 50-100 iterations, with individual evaluations ranging from seconds to half an hour depending on the basis set size. The optimisation procedure follows these key steps:

\begin{enumerate}
   \item \textbf{Data Collection}: Gather reference data from experimental measurements or accurate \textit{ab initio} calculations. This data must include the first few low-lying energy levels.
   \item \textbf{Initial Calculations}: Perform initial calculations of energy levels using a restricted basis set. These calculations provide a preliminary set of results based on the initial FMC.
   \item \textbf{Evaluation of Results}: Compare the calculated energy levels with the reference data. The evaluation is performed using a loss function, such as WRMSD, which quantifies the difference between the calculated and reference levels.
   \item \textbf{Update Surrogate Model}: Based on the evaluation, update the surrogate model that approximates the objective function. This model guides the optimisation process by predicting the impact of changes in the FMC on the calculated energy levels.
   \item \textbf{Recalculate FMC}: Use the updated surrogate model to calculate a new FMC that minimises the difference between the calculated and reference energy levels.
   \item \textbf{Convergence Check}: Convergence is verified by evaluating whether changes in the FMC lead to a significant improvement in the calculated energy levels. If convergence is not achieved, repeat the process from the evaluation step.
   \item \textbf{Final Calculation}: Once convergence is achieved, perform a full calculation using the optimised FMC and an extended basis set. This final step ensures the convergence of the calculation for a higher number of CSFs, without compromising the computational time in the previous steps.
\end{enumerate}

This optimization methodology is general and can be adapted to other atomic structure codes that rely on local central potentials or scaling parameters that require user input. This includes codes such as \HULLAC\, and \texttt{AUTOSTRUCTURE}, where scaling parameters are used. While our specific implementation targets the fictitious mean configuration in \FAC, the core principles, employing Bayesian optimization to minimize the discrepancy between calculated and reference energy levels, could be applied to any code where the potential or parameters can be systematically adjusted to improve agreement with reference data. The effectiveness of this approach would be greatest in codes where the calculated energy levels are sensitive to user-defined input parameters that affect the entire calculation consistently.The systematic evaluation of the predicted atomic data' sensitivity to the optimised mean local potential provides valuable insights into the reliability and precision of the results.

\section{Benchmark of optimisation procedure}

In this section, we present the benchmarking of our atomic structure calculations using \FAC\,, focussing on the energy levels and electric dipole (E1) transitions for selected elements. The elements chosen for this study are \ion{Au}{ii}, \ion{Pt}{ii}, and lanthanides \ion{Pr}{ii}, \ion{Pr}{iii}, \ion{Er}{ii}, \ion{Er}{iii}. For each element, we compare our calculated data with reference data from the NIST Atomic Spectra Database (ASD)\cite{Kramida2021w} and other recent experimental and theoretical studies. The results are presented, when appropriate with energy level labels using both LS and $jj$ coupling schemes. In the $jj$ coupling notation, the label is denoted as $(j_1,j_2,\dots)_J$, where $j$ are the total angular momenta of individual shells and $J$ is the total angular momentum of the configuration. While \FAC\, naturally produces results in $jj$ coupling, we employ the \texttt{JJ2LSJ} program~\cite{atoms5010006} to obtain the LS coupling labels. The identification of calculated levels with NIST experimental data was performed by matching total angular momentum $J$ and parity quantum numbers, and using the energy ordering within each $J$-parity group, similarly to what has been done in previous work \cite{Flors2023a}.

\subsection{Au II and Pt II}
\begin{figure*}[ht]
\centering
\includegraphics[width=\linewidth]{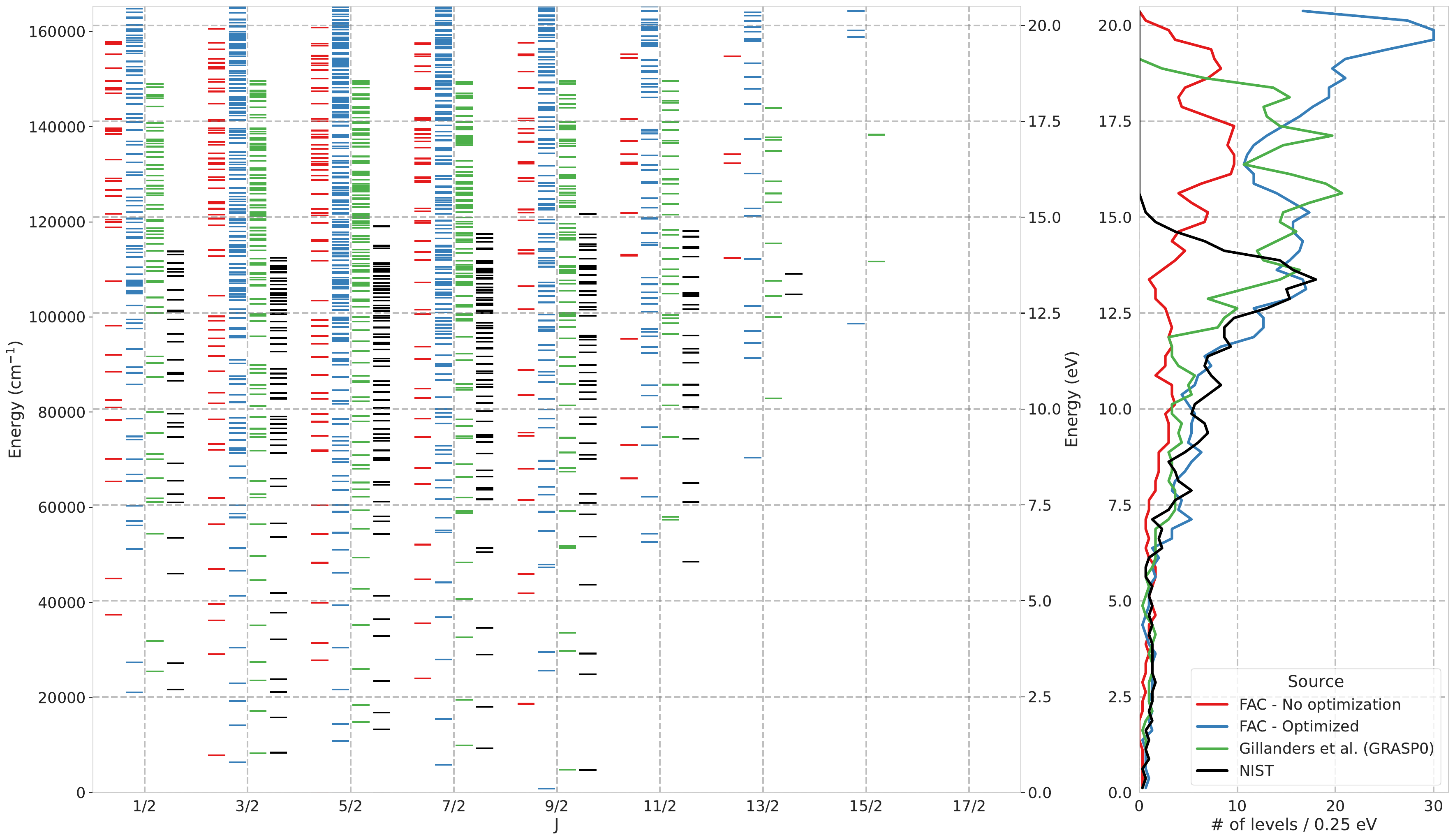}
\caption{Energy levels and level density of \ion{Pt}{ii}. Black horizontal lines show the data from the NIST ASD \cite{Kramida2021w}. Coloured horizontal lines show the calculated data using \FAC\, with (blue) and without (red) the optimisation procedure, and using \texttt{GRASP0} (green), which was used in \cite{Gillanders2021c}.}
\label{levels pt}
\end{figure*}

\begin{figure*}[ht]
\centering
\includegraphics[width=\linewidth]{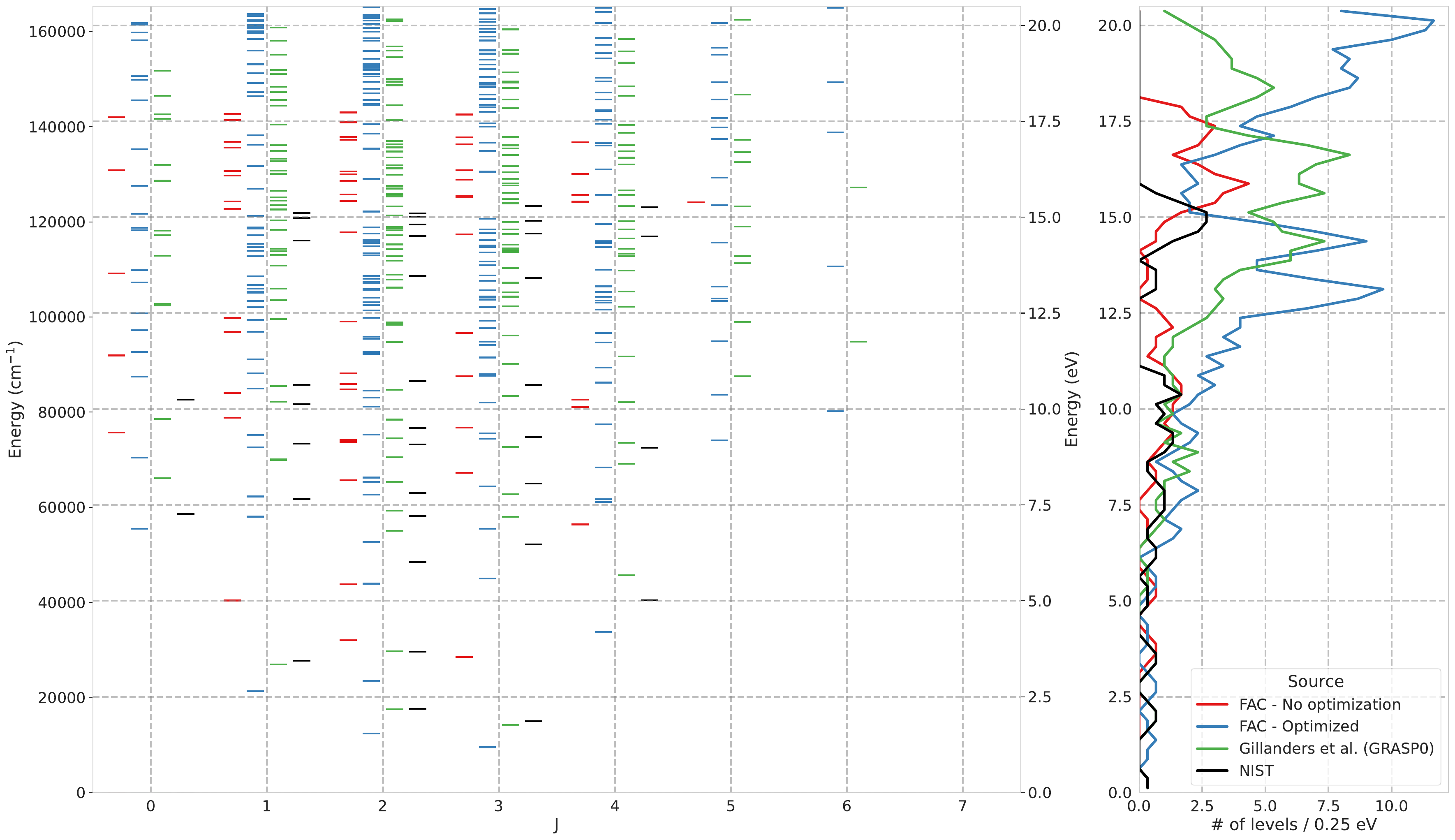}
\caption{Energy levels and level density of \ion{Au}{ii}. Black horizontal lines show the data from the NIST ASD \cite{Kramida2021w}. Coloured horizontal lines show the calculated data using \FAC\, with (blue) and without (red) the optimisation procedure, and using \texttt{GRASP0} (green), which was used in \cite{Gillanders2021c}.}
\label{levels au}
\end{figure*}

\begin{figure}[ht]
\centering
\includegraphics[width=\linewidth]{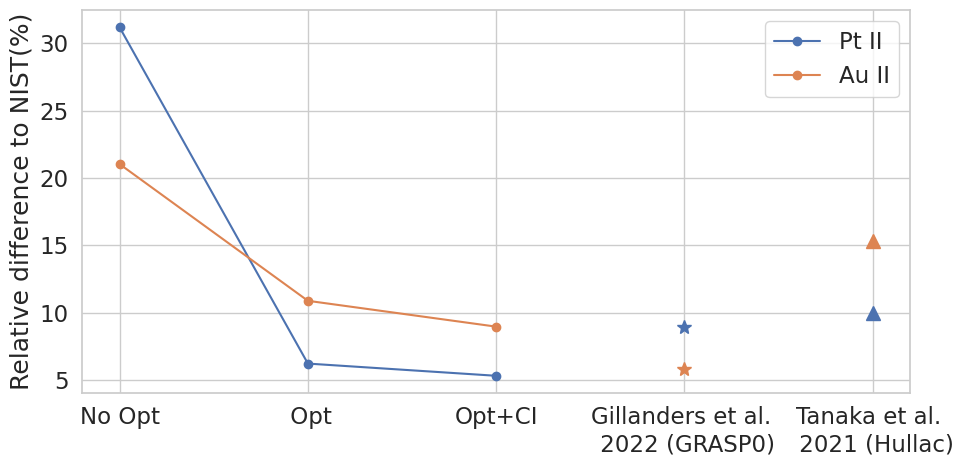}
\caption{Average relative difference to the data available in the NIST ASD \cite{Kramida2021w} for all identified levels of \ion{Pt}{ii} and \ion{Au}{ii}  for the different models computed with \FAC.``No Opt.'' corresponds to an 8 configuration model that does not use the \texttt{FAC} potential optimisation, contrarily to the other two models. The results are compared with the results from \texttt{GRASP0}\cite{Gillanders2021c} and \HULLAC \cite{Tanaka2020t, KatoOthers}, available in the literature.}
\label{fig:levels pt au}
\end{figure}

The motivation for studying \ion{Au}{ii} and \ion{Pt}{ii} arises from their potential relevance in kilonova spectra, particularly the kilonova associated with GW170817. The presence of these elements can provide critical insights into the conditions and processes occurring in outflows from neutron star mergers. Additionally, benchmarking the optimisation procedure against non-rare earth elements like Au and Pt validates our methods in a broader context.

\subsubsection{Available data}

The data available in the NIST ASD served as the reference dataset for both \ion{Au}{ii} and \ion{Pt}{ii}. For \ion{Au}{ii}, the dataset includes energy levels and transition probabilities compiled from Moore (1971)\cite{Moore1971v}, Rosberg and Wyart (1997) \cite{Rosberg1997z}, and Sansonetti and Martin (2005) \cite{JESansonettiWCMartinandSLYoung2005t}. Moore provides a compilation of atomic energy levels derived from optical spectra, including detailed information on the energy levels of \ion{Au}{ii} obtained through spectroscopic observations and measurements. Rosberg and Wyart conducted a study on the spectrum of \ion{Au}{ii} using Fourier transform spectroscopy and photographic spectrograms. Their research identified 75 new energy levels for \ion{Au}{ii}. This data was then consolidated in Sansonetti and Martin's handbook of basic atomic spectroscopic data, which is available through the NIST database.

For \ion{Pt}{ii}, extensive data is available from several key studies. Reader et al. (1988) \cite{Reader1988q} provided accurate energy levels for singly ionised platinum (\ion{Pt}{ii}) by observing the spectrum with hollow-cathode lamps and using a Fourier-transform spectrometer. Their work resulted in the measurement of 558 lines and the determination of accurate values for 28 even and 72 odd energy levels. Blaise and Wyart (1992)\cite{Blaise1992n} further contributed to the understanding of \ion{Pt}{ii} energy levels by compiling extensive experimental data, which included the use of both classical emission spectroscopy and Fourier-transform spectroscopy. They presented revised energy levels, new classifications, and improved wavelengths for numerous transitions, significantly enhancing the accuracy and consistency of the dataset. Furthermore, Wyart, Blaise, and Joshi (1995)\cite{Wyart1995d} conducted a theoretical study of the odd parity levels and transition probabilities in \ion{Pt}{ii}. They employed Hartree-Fock relativistic calculations with configuration interaction to predict energy levels and transition probabilities, providing essential theoretical insights to complement the experimental findings and addressing discrepancies in earlier experimental data.

\subsubsection{Computational procedure}

The first stage involved determining the optimal FMC using a reference dataset of accurately determined low-lying levels from the NIST ASD. The basis space for \ion{Au}{ii} and \ion{Pt}{ii} was then expanded to include all major configurations up to the ionisation energy.

For both ions, the configurations employed in the calculations were generated using single (S) and double (D) excitations from the ground state of each respective ion, extending up to $\{8s, 8p, 7d, 6f, 5g\}$. \FAC\, was used to perform the calculations both with and without FMC optimisation to assess the impact of the optimisation process on the accuracy of the results. Comparison with experimental data was performed by calculating the relative difference in excitation energies:
\begin{equation}
    \Delta E_{\text{rel}} = \frac{E_{\text{FAC}} - E_{\text{NIST}}}{E_{\text{NIST}}}
\end{equation}
where $E_{\text{FAC}}$ and $E_{\text{NIST}}$ are the excitation energies from our calculations and the NIST database, respectively.

Core-core and core-valence correlations were not included due to computational constraints. The impact of the optimisation was evaluated through direct comparison with the available experimental data.

\subsubsection{Energy Levels}

\Cref{levels pt,levels au} provide a comprehensive comparison of the energy levels for each angular momentum, $J$, for \ion{Pt}{ii} and \ion{Au}{ii}, respectively. The benchmark data from NIST ASD are compared with the results of our calculations using \FAC\, as well as the energy levels reported by Gillanders et al. using the \texttt{GRASP0} code \cite{Gillanders2021c}. In particular, the comparison is done for each $2J$ value and is not divided by parity, as the data obtained from Gillanders et al. did not include information on configurations or parity. This comparison highlights the effectiveness of the FMC optimisation in aligning our calculated energy levels with the NIST reference data, demonstrating significant improvements over the non-optimised results. For the non-optimised model, the electron occupancy number FMC was taken from an average using the ground and first configuration in energy, this is $5d^9$ and $5d^8\, 6s$ for \ion{Pt}{ii} and  $5d^{10}$ and $5d^9\, 6s$ for \ion{Au}{ii}, applying the standard semi-empirical correction built into the code. In particular, for the lowest levels, the optimised results show a typical differences in excitation energy of less than 10\% compared to the experimental data from NIST, in line with the results obtained for \texttt{GRASP0} for the singly-ionised species. These results highlight the impact of the optimisation, particularly for low-lying levels with excitation energies below 7.5 eV, where large differences are found compared to the \textit{default} \FAC\, calculation. It is worth noting that the agreement between optimized calculations and NIST data for higher-lying states (above 7.5 eV) is better for \ion{Pt}{ii} than for \ion{Au}{ii}. This difference is partially attributable to the greater abundance of experimental data for \ion{Pt}{ii} in the NIST ASD, as well as the more complex electronic structure of \ion{Au}{ii} with its filled $5d$ shell, which presents greater challenges for optimization due to stronger configuration mixing in higher-lying states.

\Cref{fig:levels pt au} illustrates the average relative difference to the data available in the NIST ASD for all identified levels of \ion{Pt}{ii} and \ion{Au}{ii} for different models computed with the \FAC\,. The ``No Opt.'' model corresponds to an 8-configuration model that does not use the \FAC\, potential optimisation, whereas the other models incorporate the optimisation. The impact of the optimisation is significant, resulting in a much greater improvement which was not achievable just by the inclusion of a larger CI basis. The results of the optimised models show a closer alignment to the NIST reference data, demonstrating similar accuracy to the \texttt{GRASP0} \cite{Gillanders2021c}  results and better energy than the \texttt{HULLAC} results \cite{Tanaka2020t, KatoOthers}.

\subsection{Pr II and Pr III}

\begin{figure*}[htb]
\centering
\includegraphics[width=\linewidth]{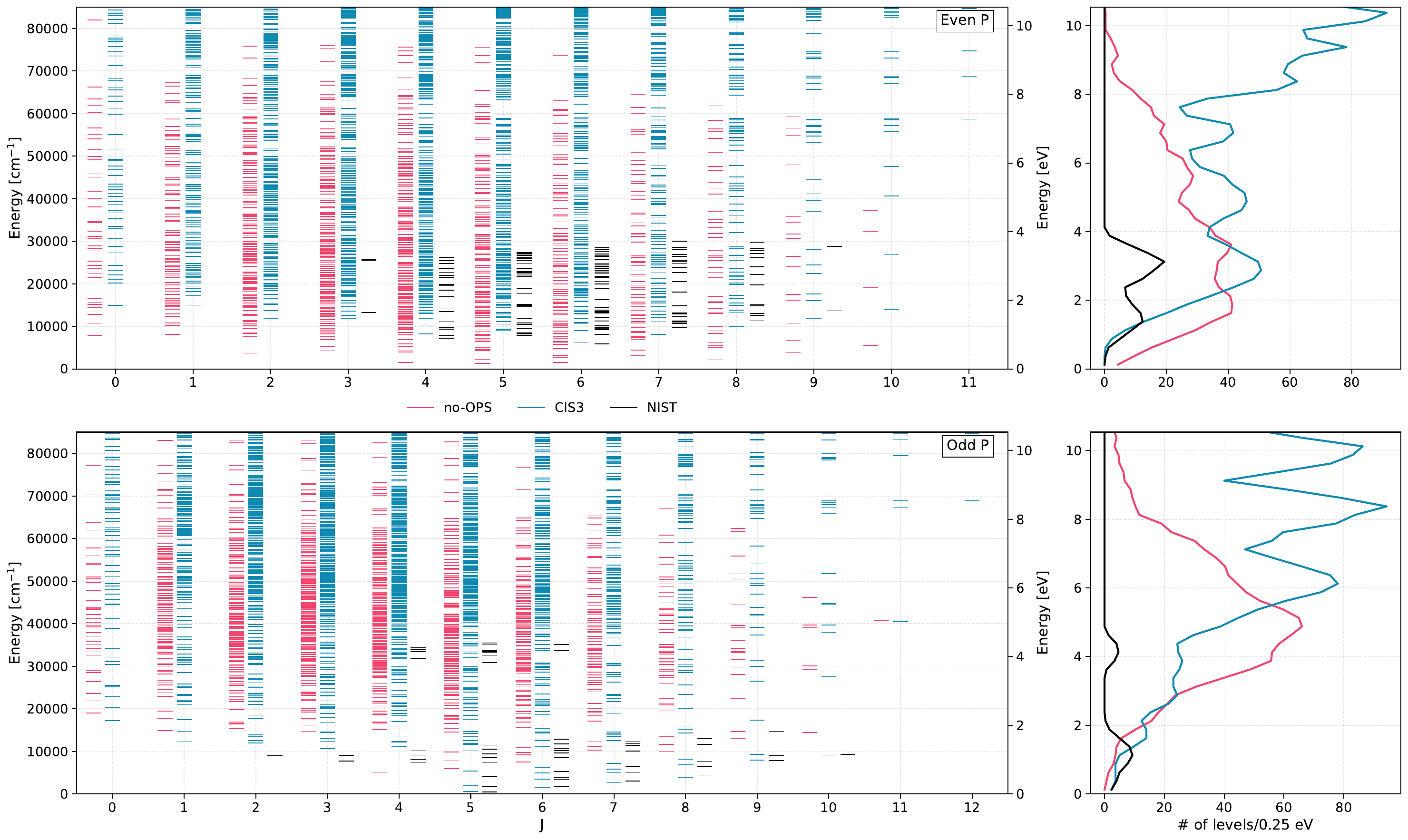}
\caption{Energy levels and level density of\ion{Pr}{ii}. Results for a calculation without optimization of the FMC (using a set of configurations equivalent to OPS) are shown alongside the results for our largest calculation, including the optimization. Black horizontal lines show the data from the NIST ASD \cite{Kramida2021w}.}
\label{fig:levels prII}
\end{figure*}

\begin{figure*}[htb]
\centering
\includegraphics[width=\linewidth]{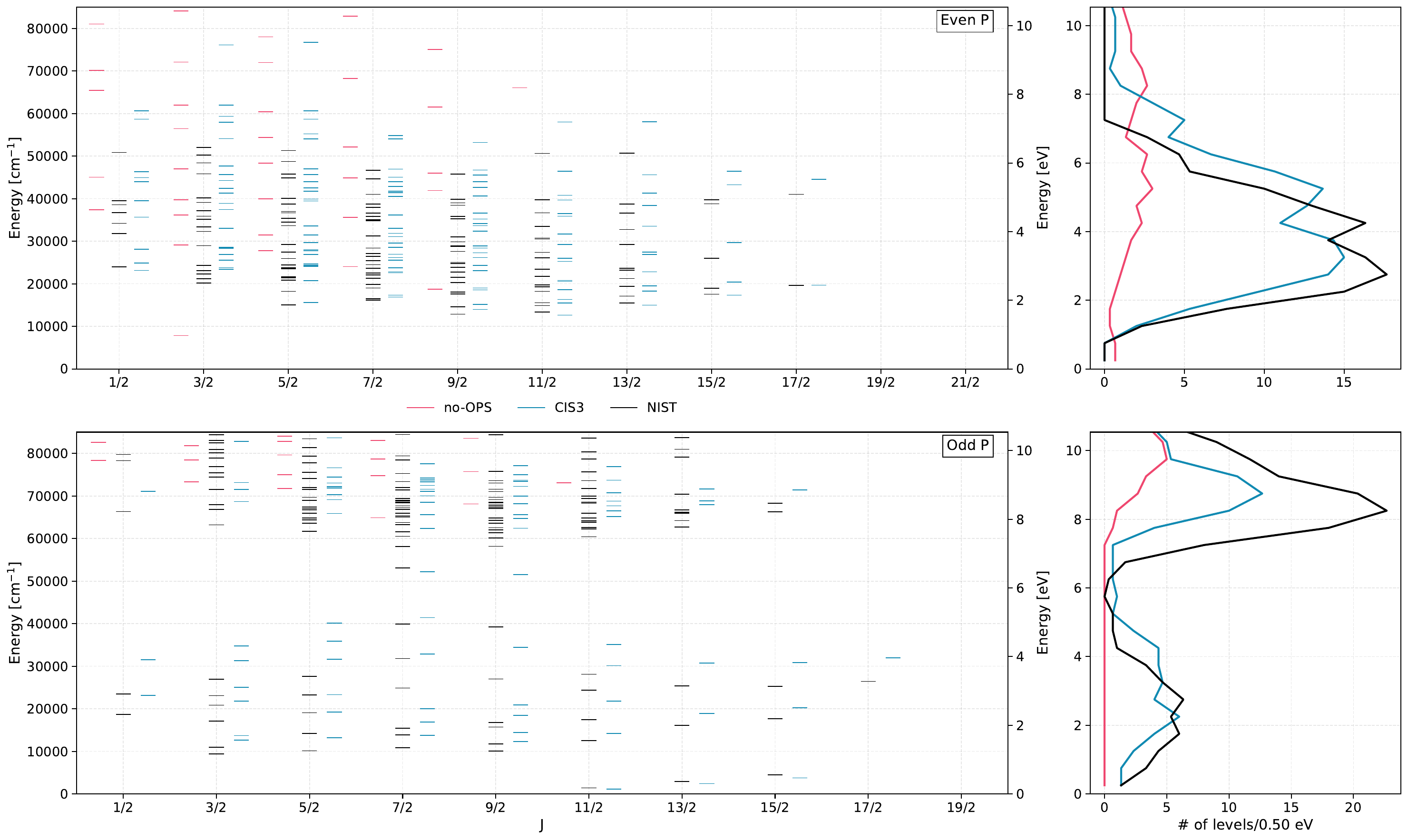}
\caption{Energy levels of and level density  \ion{Pr}{iii}. Results for a calculation without optimization of the FMC (using a set of configurations equivalent to OPS) are shown in pink alongside the results for our most complete calculation for \ion{Pr}{III}, in blue, including the optimization. Black horizontal lines show the data from the NIST ASD \cite{Kramida2021w}.}
\label{fig:levels PrIII}
\end{figure*}

\begin{table*}[ht]
\centering
\caption{Energy levels (in cm$^{-1}$) for the first 20 levels of \ion{Pr}{ii}. The columns show: electronic configuration (Config.), level labels in LS and $jj$ coupling, calculated energies without optimization ($E_{\text{no-OPS}}$), with optimization space only ($E_{\text{OPS}}$), and with increasing configuration interaction spaces ($E_{\text{CIS1}}$, $E_{\text{CIS2}}$, $E_{\text{CIS3}}$). The experimental values from NIST ($E_{\text{NIST}}$) and relative differences between CIS3 and experimental values ($\Delta_{\text{CIS3}}$) are given in the last two columns.}
\begin{tabular}{lcccccccccc}
\hline\hline
Config. & LSJ & $jj$
& $E_{\text{no-OPS}}$ & $E_{\text{OPS}}$ & $E_{\text{CIS1}}$ & $E_{\text{CIS2}}$ & $E_{\text{CIS3}}$ & $E_{\text{NIST}}$ &  $\Delta_{\text{CIS3}}(\%)$ \\
\hline
$4f_{3} \, 6s$ & ${}^{5}I_{4}$ & $(9/2,1/2)_{4}$ & 5080.19 & 0.00 & 0.00 & 0.00 & 0.00 & 0.00 & --- \\
$4f_{3} \, 6s$ & ${}^{5}I_{5}$ & $(9/2,1/2)_{5}$ & 5925.82 & 591.86 & 589.12 & 592.23 & 566.52 & 441.95 & 28.19\% \\
$4f_{3} \, 6s$ & ${}^{5}I_{6}$ & $(11/2,1/2)_{6}$ & 7532.35 & 1746.43 & 1810.25 & 1707.48 & 1512.49 & 1649.01 & 8.28\% \\
$4f_{3} \, 6s$ & ${}^{3}I_{5}$ & $(11/2,1/2)_{5}$ & 7786.34 & 2012.86 & 1998.63 & 2007.36 & 1942.94 & 1743.72 & 11.42\% \\
$4f_{3} \, 6s$ & ${}^{5}I_{7}$ & $(13/2,1/2)_{7}$ & 8902.52 & 3554.95 & 3114.87 & 3073.74 & 2634.58 & 2998.36 & 12.13\% \\
$4f_{3} \, 6s$ & ${}^{3}I_{6}$ & $(13/2,1/2)_{6}$ & 9726.35 & 3699.41 & 3778.21 & 3532.30 & 3445.13 & 3403.21 & 1.23\% \\
$4f_{3} \, 5d$ & ${}^{5}L_{6}$ & $(9/2,3/2)_{6}$ & 9179.41 & 5158.62 & 5115.33 & 5052.51 & 4915.09 & 3893.46 & 26.24\% \\
$4f_{3} \, 5d$ & ${}^{5}K_{5}$ & $(9/2,3/2)_{5}$ & 17789.41 & 5595.43 & 5590.27 & 5649.02 & 5393.42 & 4097.60 & 31.62\% \\
$4f_{3} \, 6s$ & ${}^{5}I_{8}$ & $(15/2,1/2)_{8}$ & 10004.31 & 4669.32 & 4744.13 & 4675.17 & 3883.28 & 4437.15 & 12.48\% \\
$4f_{3} \, 6s$ & ${}^{3}I_{7}$ & $(15/2,1/2)_{7}$ & 11159.94 & 5776.56 & 5652.82 & 5217.45 & 4995.07 & 5079.35 & 1.66\% \\
$4f_{3} \, 5d$ & ${}^{5}L_{7}$ & $(11/2,3/2)_{7}$ & 18355.79 & 6083.82 & 5912.63 & 6010.73 & 5814.83 & 5108.40 & 13.83\% \\
$4f_{3} \, 5d$ & ${}^{5}K_{6}$ & $(11/2,3/2)_{6}$ & 16917.59 & 6489.45 & 6517.05 & 6306.81 & 6217.95 & 5226.52 & 18.97\% \\
$4f_{2} \, 5d_{2}$ & ${}^{5}L_{6}$ & $(4,6)_{6}$ & 5755.23 & 6569.53 & 6553.05 & 6402.30 & 6280.25 & 5854.61 & 7.27\% \\
$4f_{3} \, 5d$ & ${}^{5}L_{7}$ & $(11/2,5/2)_{7}$ & 10297.00 & 7306.01 & 7416.36 & 7498.47 & 7142.73 & 6413.93 & 11.36\% \\
$4f_{3} \, 5d$ & ${}^{5}K_{8}$ & $(11/2,5/2)_{8}$ & 21291.86 & 6960.23 & 7027.83 & 6999.42 & 6820.46 & 6417.83 & 6.27\% \\
$4f_{2} \, 5d_{2}$ & ${}^{3}I_{4}$ & $(5,7)_{4}$ & 7483.54 & 8491.31 & 8251.76 & 8568.32 & 8238.45 & 7227.99 & 13.98\% \\
$4f_{3} \, 5d$ & ${}^{5}G_{5}$ & $(9/2,5/2)_{5}$ & 16576.42 & 10452.05 & 10587.04 & 10426.38 & 10114.77 & 7438.23 & 35.98\% \\
$4f_{3} \, 5d$ & ${}^{5}I_{4}$ & $(9/2,5/2)_{4}$ & 15049.01 & 11302.87 & 10910.87 & 11116.26 & 10788.88 & 7446.43 & 44.89\% \\
$4f_{3} \, 5d$ & ${}^{5}K_{8}$ & $(13/2,5/2)_{8}$ & 13454.15 & 8410.92 & 8765.74 & 8267.51 & 8158.91 & 7659.76 & 6.52\% \\
$4f_{3} \, 5d$ & ${}^{3}H_{3}$ & $(9/2,5/2)_{3}$ & 14705.94 & 10817.93 & 10830.89 & 11031.73 & 10595.78 & 7744.27 & 36.82\% \\
\hline\hline
\end{tabular}
\label{tab:energy_levels_prii}
\end{table*}

\begin{table*}[ht]
\centering
\caption{Energy levels (in cm$^{-1}$) for the first 20 levels of \ion{Pr}{iii}. The columns show: electronic configuration (Config.), level labels in LS and $jj$ coupling,, calculated energies without optimization ($E_{\text{no-OPS}}$), with optimization space only ($E_{\text{OPS}}$), and with increasing configuration interaction spaces ($E_{\text{CIS1}}$, $E_{\text{CIS2}}$, $E_{\text{CIS3}}$). The experimental values from NIST ($E_{\text{NIST}}$) and relative differences between CIS3 and experimental values ($\Delta_{\text{CIS3}}$) are given in the last two columns.}
\begin{tabular}{lccccccccc}
\hline\hline
Config. & LSJ & $jj$ & $E_{\text{no-OPS}}$ & $E_{\text{OPS}}$ & $E_{\text{CIS1}}$ & $E_{\text{CIS2}}$ & $E_{\text{CIS3}}$ & $E_{\text{NIST}}$ & $\Delta_{\text{CIS3}}(\%)$\\
\hline
$4f_{3}$ & ${}^{4}I_{9/2}$ & $(9/2)_{9/2}$ & $0.00$ & $0.00$ & $0.00$ & $0.00$ & $0.00$ & $0.00$ & --- \\
$4f_{3}$ & ${}^{4}I_{11/2}$ & $(11/2)_{11/2}$ & $1326.93$ & $1592.85$ & $1352.95$ & $1357.94$ & $1331.27$ & $1398.34$ & 4.80\% \\
$4f_{3}$ & ${}^{4}I_{13/2}$ & $(13/2)_{13/2}$ & $2427.11$ & $3054.46$ & $2812.87$ & $2767.48$ & $2601.63$ & $2893.14$ & 10.08\% \\
$4f_{3}$ & ${}^{4}I_{15/2}$ & $(15/2)_{15/2}$ & $3247.11$ & $4713.66$ & $5393.79$ & $4112.37$ & $4983.26$ & $4453.76$ & 11.89\% \\
$4f_{3}$ & ${}^{4}F_{3/2}$ & $(3/2)_{3/2}$ & $11040.02$ & $8700.01$ & $9100.72$ & $9704.16$ & $10040.01$ & $9370.66$ & 7.14\% \\
$4f_{3}$ & ${}^{4}S_{5/2}$ & $(5/2)_{5/2}$ & $11715.30$ & $9501.93$ & $7914.18$ & $10433.33$ & $10355.69$ & $10138.18$ & 2.15\% \\
$4f_{3}$ & ${}^{2}H_{7/2}$ & $(7/2)_{7/2}$ & $12256.66$ & $11283.33$ & $12769.56$ & $12425.54$ & $9712.14$ & $10859.06$ & 10.56\% \\
$4f_{3}$ & ${}^{4}F_{3/2}$ & $(3/2)_{3/2}$ & $12113.48$ & $9868.64$ & $11256.59$ & $12871.12$ & $9985.74$ & $10950.24$ & 8.81\% \\
$4f_{3}$ & ${}^{2}H_{9/2}$ & $(9/2)_{9/2}$ & $12775.43$ & $12744.02$ & $12852.55$ & $12138.72$ & $11314.66$ & $11761.69$ & 3.80\% \\
$4f_{3}$ & ${}^{2}G_{11/2}$ & $(11/2)_{11/2}$ & $12417.17$ & $10997.75$ & $14184.62$ & $11278.18$ & $13289.73$ & $12494.63$ & 6.36\% \\
$4f_{2} \, 5d$ & ${}^{4}D_{9/2}$ & $(4,3/2)_{9/2}$ & $14160.01$ & $11819.85$ & $15749.80$ & $11285.49$ & $13874.46$ & $12846.66$ & 8.00\% \\
$4f_{2} \, 5d$ & ${}^{2}L_{11/2}$ & $(4,3/2)_{11/2}$ & $12614.12$ & $14073.38$ & $11151.44$ & $12656.52$ & $13740.95$ & $13352.10$ & 2.91\% \\
$4f_{3}$ & ${}^{2}D_{7/2}$ & $(7/2)_{7/2}$ & $14885.84$ & $12501.43$ & $15249.68$ & $13407.63$ & $12854.25$ & $13887.60$ & 7.44\% \\
$4f_{3}$ & ${}^{4}D_{5/2}$ & $(5/2)_{5/2}$ & $16882.02$ & $15466.66$ & $12763.88$ & $15297.72$ & $13752.29$ & $14187.35$ & 3.07\% \\
$4f_{2} \, 5d$ & ${}^{2}I_{9/2}$ & $(4,5/2)_{9/2}$ & $15542.58$ & $15029.91$ & $11954.67$ & $11907.78$ & $15121.47$ & $14558.82$ & 3.86\% \\
$4f_{2} \, 5d$ & ${}^{2}F_{11/2}$ & $(5,11/2)_{11/2}$ & $14874.72$ & $16998.92$ & $14070.40$ & $16619.43$ & $15099.27$ & $14859.96$ & 1.61\% \\
$4f_{2} \, 5d$ & ${}^{2}K_{5/2}$ & $(4,5/2)_{5/2}$ & $15288.53$ & $17417.69$ & $12580.62$ & $17779.54$ & $15474.70$ & $15045.80$ & 2.85\% \\
$4f_{3}$ & ${}^{2}H_{7/2}$ & $(7/2)_{7/2}$ & $17735.69$ & $14016.27$ & $18781.51$ & $15082.44$ & $13650.54$ & $15443.48$ & 11.61\% \\
$4f_{2} \, 5d$ & ${}^{2}L_{13/2}$ & $(5,13/2)_{13/2}$ & $14299.77$ & $13519.77$ & $13758.01$ & $16407.71$ & $14450.47$ & $15454.16$ & 6.49\% \\
$4f_{2} \, 5d$ & ${}^{2}D_{11/2}$ & $(4,11/2)_{11/2}$ & $16863.21$ & $17168.58$ & $12301.18$ & $13995.94$ & $17088.29$ & $15525.50$ & 10.07\% \\
\hline\hline
\end{tabular}
\label{tab:energy_levels_priii}
\end{table*}

\subsubsection{Available data}

A first study on the spectrum of singly-ionised praseodymium was conducted by Rosen et al. (1941) \cite{Rosen1941n}, which determined energies, $g$ and $J$ values for 74 levels from resolved Zeeman patterns of 141 lines in the UV and optical range (2400 to 7100 \AA). Ginibre (1989) \cite{Ginibre1989s} studied 105 odd and 187 even experimental energy levels using Fourier transform (FT) spectroscopy over the range of 2783–25 000 cm$^{-1}$. Furthermore, Wyart et al. (1974) \cite{Wyart1974h} studied a significant number of levels corresponding to the $4f^N (5d + 6s)$ configurations for multiple lanthanides, including \ion{Pr}{ii} and \ion{Pr}{iii}. Later, LS-coupling labelling was assigned to many of the identified levels using a semi-empirical fitting approach by Ginibre (1990) \cite{Ginibre1990e}. Later experimental measurements were conducted by Ivarsson et al. (2001) \cite{Ivarsson2001j}, which adjusted the energy of 39 levels using FT spectroscopy in the 2800–8000 \AA range. Furmann et al. (2005, 2007) \cite{Furmann2005n, Furmann2007i} examined 31 odd and 14 even levels using laser-induced fluorescence (LIF) spectroscopy in a hollow cathode discharge lamp. More recently, Akhtar \& Windholz (2012) \cite{Akhtar2012k} re-evaluated the energy values for 227 levels (74 odd and 153 even parity) and the hyperfine structures of 477 transitions in the 3260–11700 Å range, correcting the energy levels from Ginibre and Ivarsson et al. All these levels were measured or reanalysed with high precision. The data from these studies are summarised in \cite[Martin et al. (1978)]{Martin1978o}, and this work is included in the NIST ASD.

\subsubsection{Computational procedure}

The first stage involved computing the FMC used to determine the optimal potential. This was done using a reference dataset that included at least a few accurately determined low-lying levels. The data from the NIST ASD was used as the main reference for building this dataset.

The basis space for \ion{Pr}{ii} and \ion{Pr}{iii} was expanded to include all major configurations up to the ionisation energy. For \ion{Pr}{ii} and \ion{Pr}{iii}, it was found that considering single and double SD excitations from the ground configuration ($4f^3 \,6s$ for \ion{Pr}{ii} and $4f^3$ for \ion{Pr}{iii}) up to {$7s$, $7p$, $7d$, $5f$, $5g$} effectively covered the most relevant configurations that contribute directly to opacity \cite{Flors2023a}. These configurations were used in the optimisation procedure in order to determine the optimal FMC for these ions. Hence, we label this space as Optimisation Space (OPS). To further enhance the convergence of the results, additional configurations were included based on SD excitations extending up to principal quantum number $n=10$) angular momentum $\ell=5$. The space was increased in multiple steps from a base ground configuration (GC), with each space used labelled as a different CI space (CIS). The scheme for the full set of calculations was as follows.

\begin{itemize}
\item OPS =  GC + SD\{$7s$, $7p$, $7d$, $5f$, $5g$\} 
\item CIS1 = OPS + SD\{$8s$, $8p$, $8d$, $6f$, $5g$\}
\item CIS2 = CIS1 + SD\{$9s$, $9p$, $9d$, $7f$, $6g$\}
\item CIS3 = CIS2 + SD\{$10s$, $10p$, $10d$, $7f$, $6g$\}
\end{itemize}

While core-core and core-valence correlations are expected to improve the calculations for these ions \cite{Safronova2015y}, they were found to be too computationally demanding. Moreover, the marginal improvements in accuracy potentially gained from these correlations may not justify the significant increase in computational resources required.

\subsubsection{Energy Levels}

The energy levels computed for \ion{Pr}{ii} and \ion{Pr}{iii} are compared with the NIST recommended values in Figures  \cref{fig:levels prII} and  \cref{fig:levels PrIII}. The results without optimisation (using a set of configurations equivalent to OPS) show a significant deviation from the NIST values, with an average relative difference of approximately 60\% for \ion{Pr}{ii} and 20\% for \ion{Pr}{iii}. The optimised results, however, show a marked improvement, reducing the average relative difference to about 10\% for \ion{Pr}{ii} and 8\% for \ion{Pr}{iii}. The black horizontal lines show the data from the NIST ASD. This improvement underscores the efficacy of the FMC optimisation in enhancing the accuracy of our computed energy levels. Additionally, the results indicate that the optimised configurations effectively capture the electron correlation effects, which are crucial for achieving close agreement with the experimental data. 

By examining the figures, it is evident that the optimisation particularly benefits the lower energy levels, whereas the non optimised results tend to diverge more significantly from the NIST values. This trend is consistent across both \ion{Pr}{ii} and \ion{Pr}{iii}, highlighting the robustness of the optimisation approach. The figures also demonstrate that the higher energy levels, while improved, still exhibit some discrepancies. These discrepancies can be attributed to the higher energy levels being more scarce and often isolated in the NIST database (e.g., odd parity \ion{Pr}{ii}), which may or may not allow great results for these levels.

\begin{figure}[ht]
\centering
\includegraphics[width=\linewidth]{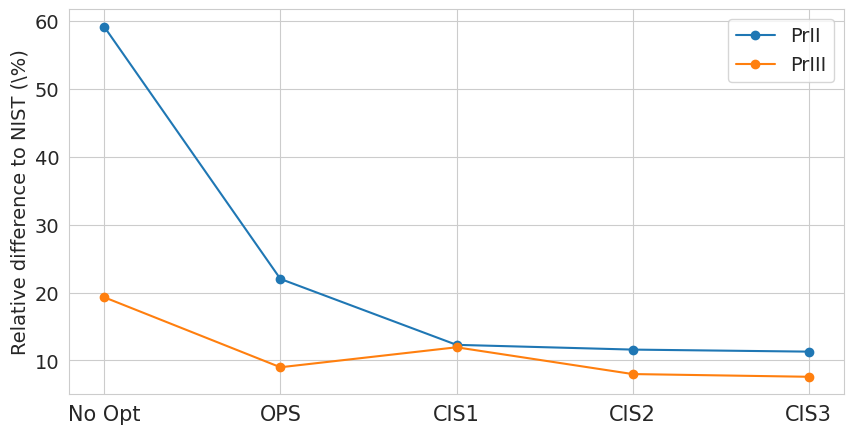}
\caption{Average relative difference to the data available in the NIST ASD \cite{Kramida2021w} for all identified levels of \ion{Pr}{ii} and \ion{Pr}{iii} for all the different models computed with \FAC. ``No Opt.'' corresponds to an OPS configuration model that does not use the \texttt{FAC} potential optimisation, contrary to the other models.}
\label{fig:levels Pr opti}
\end{figure}

The average relative difference from the NIST data for all identified levels of \ion{Pr}{ii} and \ion{Pr}{iii} for the different models calculated with \FAC\, is shown in \cref{fig:levels Pr opti}. The ``No Opt." model corresponds to the OPS configuration model that does not use the \FAC\, potential optimisation, whereas the other models incorporate this optimisation. The optimisation leads to a substantial reduction in the relative difference, far exceeding the improvements achieved by CI alone. Additionally, the convergence of results across different CI spaces (CIS1, CIS2, and CIS3) indicates that the optimisation process stabilises the calculations, ensuring reliable and consistent results. The results for the energy levels of the first 20 computed levels are shown in \cref{tab:energy_levels_prii,tab:energy_levels_priii}

\subsubsection{Transitions}
\begin{figure}
\centering
\includegraphics[width=\linewidth]{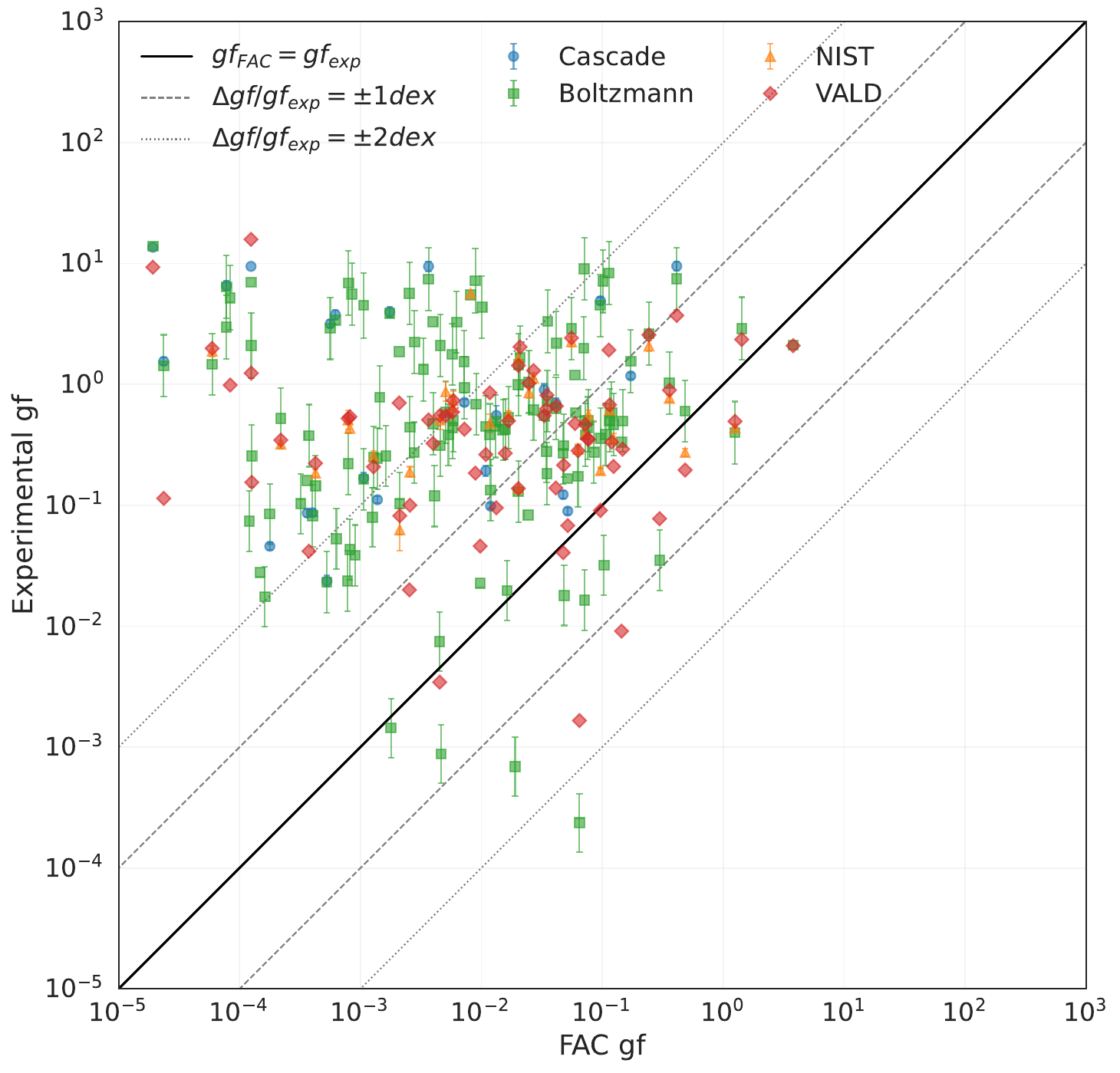}
\caption{Comparison of calculated $\log(gf)$ values from ${CIS3}$ model with experimental data for \ion{Pr}{ii}. The y-axis shows the experimental $\log(gf)$ values, while the x-axis represents the calculated values. Different colours indicate the source of the experimental data: cascade (blue circles) and Boltzmann plot method (orange triangles) available from Ferrara et al. (2024) \cite{Ferrara2023q}; VALD (red squares) and the NIST database (orange triangles).}
\label{fig:prgf}
\end{figure}
\begin{figure}
\centering
\includegraphics[width=\linewidth]{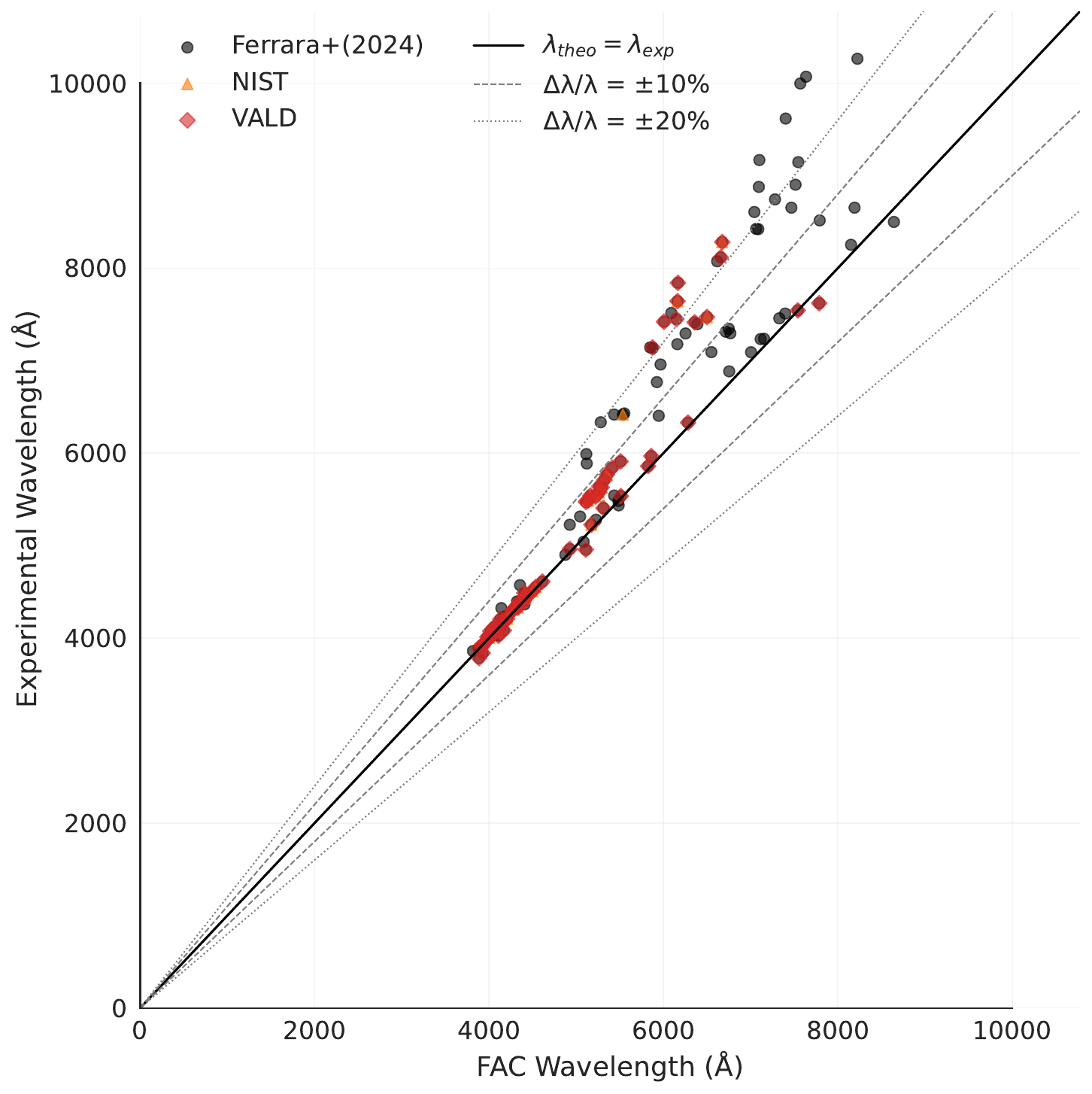}
\caption{Evaluation of calculated wavelengths from ${CIS3}$ model with experimental data for \ion{Pr}{ii}. The y-axis shows the experimental wavelengths in Angstroms, while the x-axis represents the calculated wavelengths. Grey dots represent comparison to available experimental data from Ferrara et al. (2024) \cite{Ferrara2023q}, while orange triangles and red squares compare our data with data available in VALD and NIST ASD, respectively.  The dashed lines indicate a 10\% difference from the experimental data, while the dotted lines show a 20\% difference.}
\label{fig:prwv}
\end{figure}
To evaluate the accuracy of our calculations for radiative transition probabilities and wavelengths in \ion{Pr}{ii} and \ion{Pr}{iii}, we compared our results with recent experimental data from Ferrara et al. (2024) \cite{Ferrara2023q}. Their study provides valuable spectroscopic data for several high-$Z$ elements, including praseodymium, covering the 3700-10000 \AA\ range using high-resolution \'echelle spectroscopy.
\Cref{fig:prgf} presents a comparison between the calculated $\log(gf)$ values from our {CIS3} model and the experimental data. The experimental values are derived using two methods: Boltzmann plots and cascade calculations. It's important to note that while cascade method results are generally reliable within the limits of adopted literature values, the Boltzmann method is only valid for transitions in thermodynamic equilibrium. When available, we also evaluated our data against data from the Vienna Atomic Line Database (VALD) \cite{2015PhyS...90e4005R}. As evident from \cref{fig:prgf}, there is considerable scatter in the $\log(gf)$ values, particularly for weaker transitions ($\log(gf) < -2$).  The agreement between our calculations and the experimental data improves for stronger transitions, but discrepancies persist across the entire range. Quantitatively, we observe a root mean square deviation (RMSD) of approximately 1.1 dex between our calculated $\log(gf)$ values and the experimental data. This level of disagreement is not uncommon in atomic structure calculations for complex ions like \ion{Pr}{ii} and \ion{Pr}{iii}, especially given the challenges in accurately representing the electron correlations in these systems.
In contrast, \cref{fig:prwv} shows good agreement between our calculated wavelengths and the experimental values. Most of the calculated wavelengths fall within 10\% of the experimental data, as indicated by the points between the inner dotted lines. Nearly all calculated wavelengths are within 20\% of the experimental values, falling between the dashed lines. Quantitatively, we find that approximately 80\% of our calculated wavelengths deviate less than 20\% from the experimental values and 96\% are within 10\%.
This disparity in accuracy between the radiative transition probabilities and wavelengths is a common feature in atomic structure calculations. Wavelengths, which are primarily determined by energy level differences, are often more accurately predicted than transition probabilities, which depend on the precise details of the wavefunctions.

\subsection{Er II and Er III}

\subsubsection{Available data}
Extensive research has been conducted on the energy levels and spectral characteristics of singly ionised erbium (\ion{Er}{ii}), with many studies advancing the understanding of its configurations and transitions. McNally \& Sluis (1959) \cite{McNally1959t} detailed eight low-lying levels of the $4f^{12}\, 6s$ configuration, drawing on Zeeman data from Lindner \& Davis (1958) \cite{Lindner1958h} and wavelength measurements from Gatterer \& Junkes (1945) \cite{Gatterer1945p}. These levels were further categorised into four $Jj$ terms by Judd \& Marquet (1962) \cite{Judd1962q}, who observed that the spacings between components were affected by the Coulomb interaction between the $6s$ electron and the $4f^{12}$ core.

Building on this work, Goldschmidt (1963) \cite{Goldschmidt1963b} computed all 24 levels of the $4f^{12}$ 6s configuration by fitting the $F_{2n}$ Slater parameters using the initial eight experimentally determined levels as a reference. Subsequently, Marquet \& Davis (1965) \cite{Marquet1965b} identified 10 levels of the ground configuration using wavelengths in the visible and near-ultraviolet regions. Their Zeeman structure measurements of 132 transitions enabled Sluis \& McNally (1970) \cite{VanderSluis1970v} to identify 12 of the lowest ground configuration levels, which exhibited slightly higher energies than those calculated by Goldschmidt (1963) with $E^2$ = 30. Additionally, Sluis \& McNally (1970) reported 90 odd levels, but these were excluded from comparisons due to a lack of configuration identification and $J$ assignment.

Spector (1971) \cite{Spector1971u} examined infrared photography from electrodeless discharge and sliding spark, identifying four levels of the $4f$ $5d$ configuration, which are included in the NIST database. This analysis encompassed 40 odd levels in the 25000–33000 cm$^{-1}$ energy range. The data from the aforementioned studies, including an unpublished extension of the analysis by van Kleef and Koot (1975), have been compiled and systematised in the Handbook of Basic Atomic Spectroscopic by J. E. Sansonetti and W. C. Martin (2005) \cite{JESansonettiWCMartinandSLYoung2005t} and are available in the NIST database.

In more recent work, Wyart \& Lawler (2009) \cite{Wyart2009y} expanded the spectrum classification both theoretically and experimentally. Initially, they interpreted known energy levels parametrically using Cowan’s code \cite{Cowan1981n} on two sets of interacting configurations. These predictions extended the classification of hollow-cathode FT spectra of erbium recorded at the US National Solar Observatory on Kitt Peak. Using CI, they assigned all 130 known levels of even parity to the $4f^{12}\{6s, 5d\}$ and $4f^{11}\{6s6p, 5d6p\}$ configurations, and 230 known levels of odd parity to the $4f^{12}\, 6p$ and $4f^{11}\{5d6s, 5d^2, 6s^2\}$ configurations. They also introduced 32 new energy levels and revised the $J$-values for six levels.

Research into the energy levels and spectral characteristics of doubly ionised erbium (\ion{Er}{iii}) has identified both even and odd configurations. For even configurations, the identified levels include $4f^{12}$ and $4f^{11}\,6p$, while odd configurations are $4f^{11}$6s and $4f^{11}\,5d$.

Becher (1966) \cite{J1966j} was one of the first to analyse its spectrum by identifying nine levels within the $4f^{11}(4f_{15/2})5d$, $4f^{11}(4f_{15/2})6s$, and $4f^{11}(4f_{15/2})6p$ groups. He noted the $J_1j$ coupling in these sub-configurations and assigned levels to all five $J_1j$ terms based on the $4f^{11}(4f_{15/2}$) parent level.

Most of the levels and term assignments were further detailed by Spector (1973) \cite{Spector1973q}, who independently found a more comprehensive system of excited levels and positioned them relative to three levels of the $4f^{12}$ ground configuration, including the ground level. Wyart, Blaise, and Camus (1974) \cite{Wyart1974h} studied the $(4f^{11}\,5d + 4f^{11}\,6s)$ configurations for \ion{Er}{iii}, adding two levels and suggesting the removal of two others from Spector's list. They also calculated the leading percentages in LS coupling from their paper, noting that three of these levels could not be unambiguously assigned to calculated eigenvectors.

Wyart, Koot, and van Kleef (1974) \cite{Wyart1974x} further calculated the percentages in $J_1j$ coupling for the $4f^{11}6p$ group of levels and identified two new levels of $4f^{11}\,5d$. Lastly, the data from Becher were reanalyzed by Wyart et al. (1997) \cite{Wyart1997t} increasing the number of identified levels from 45 to 115.

\subsubsection{Computational Procedure}

The computational procedure for \ion{Er}{ii} and \ion{Er}{iii} followed a similar approach to that used for \ion{Pr}{ii} and \ion{Pr}{iii}. The first stage involved determining the fictitious mean configuration (FMC) to establish the optimal potential, using a reference dataset that included several accurately determined low-lying levels. The NIST ASD served as the main reference for constructing this dataset.

The basis space for \ion{Er}{ii} and \ion{Er}{iii} was expanded to include all major configurations up to the ionisation energy. SD excitations from the ground configuration ($4f^{12}\,6s$ for \ion{Er}{ii} and $4f^{12}$ for \ion{Er}{iii}) up to $\{7s, 7p, 7d, 5f, 5g\}$ were considered. This space was once again designated as the Optimisation Space (OPS) in the optimisation procedure to determine the optimal FMC.

Due to computational constraints, excitations were included up to CIS2. No core-core or core-valence correlations were included. The configuration space was expanded as follows:

\begin{itemize}
    \item OPS = GC + SD$\{7s, 7p, 7d, 5f, 5g\}$
    \item CIS1 = OPS + SD$\{8s,\, 8p,\, 8d,\, 6f,\, 5g\}$
    \item CIS2 = CIS1 + SD$\{9s,\, 9p,\, 9d,\, 7f,\, 6g\}$
\end{itemize}

\subsubsection{Energy Levels}

\begin{figure*}[ht]
\centering
\includegraphics[width=\linewidth]{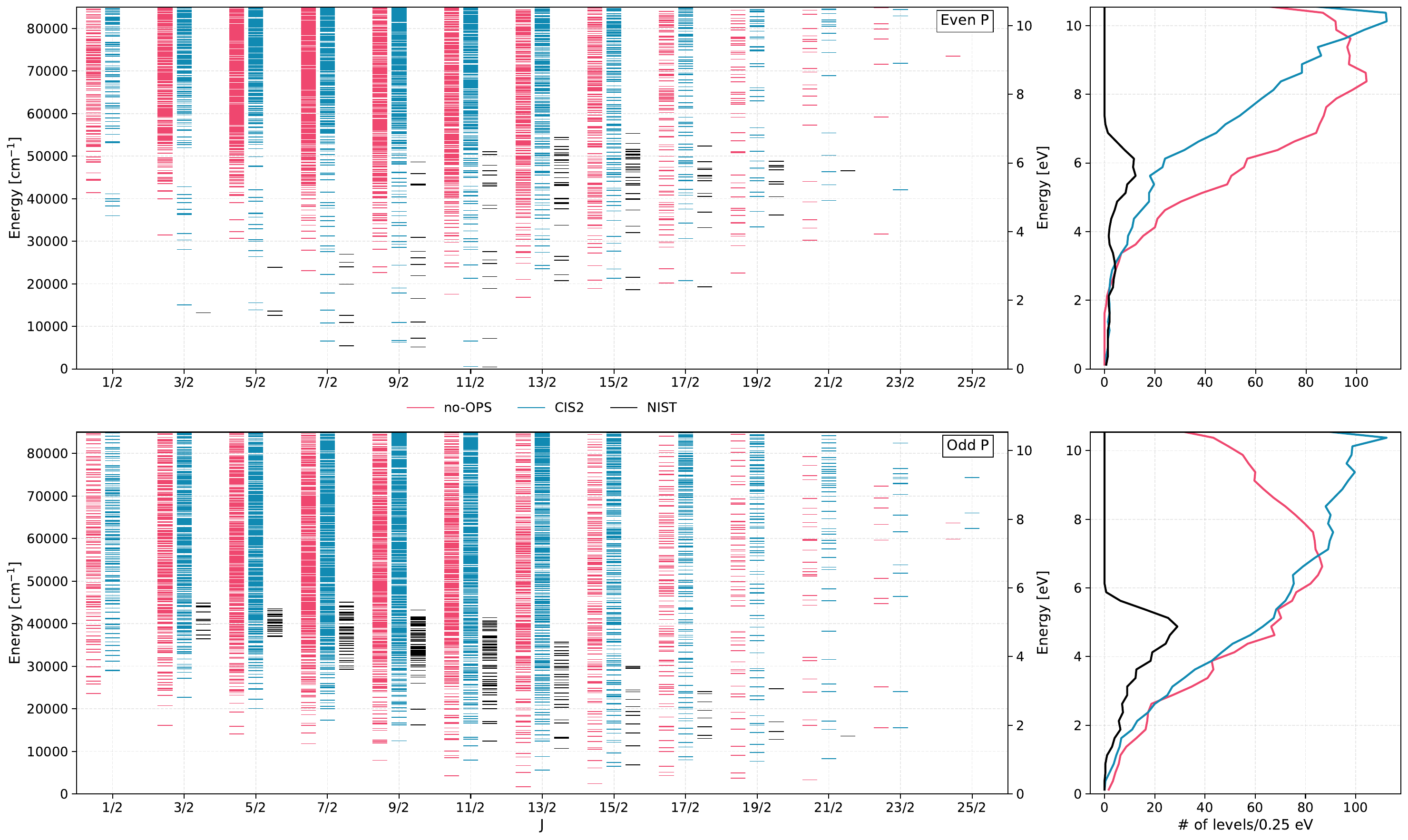}
\caption{Energy levels and level density of \ion{Er}{ii}. Pink lines show results for a calculation without optimisation of the FMC (using a set of configurations equivalent to OPS) while blue lines show the results for our largest calculation, including the optimisation. Black horizontal lines show the data from the NIST ASD \cite{Kramida2021w}}
\label{fig:levels ErII}
\end{figure*}

\begin{figure*}[ht]
\centering
\includegraphics[width=\linewidth]{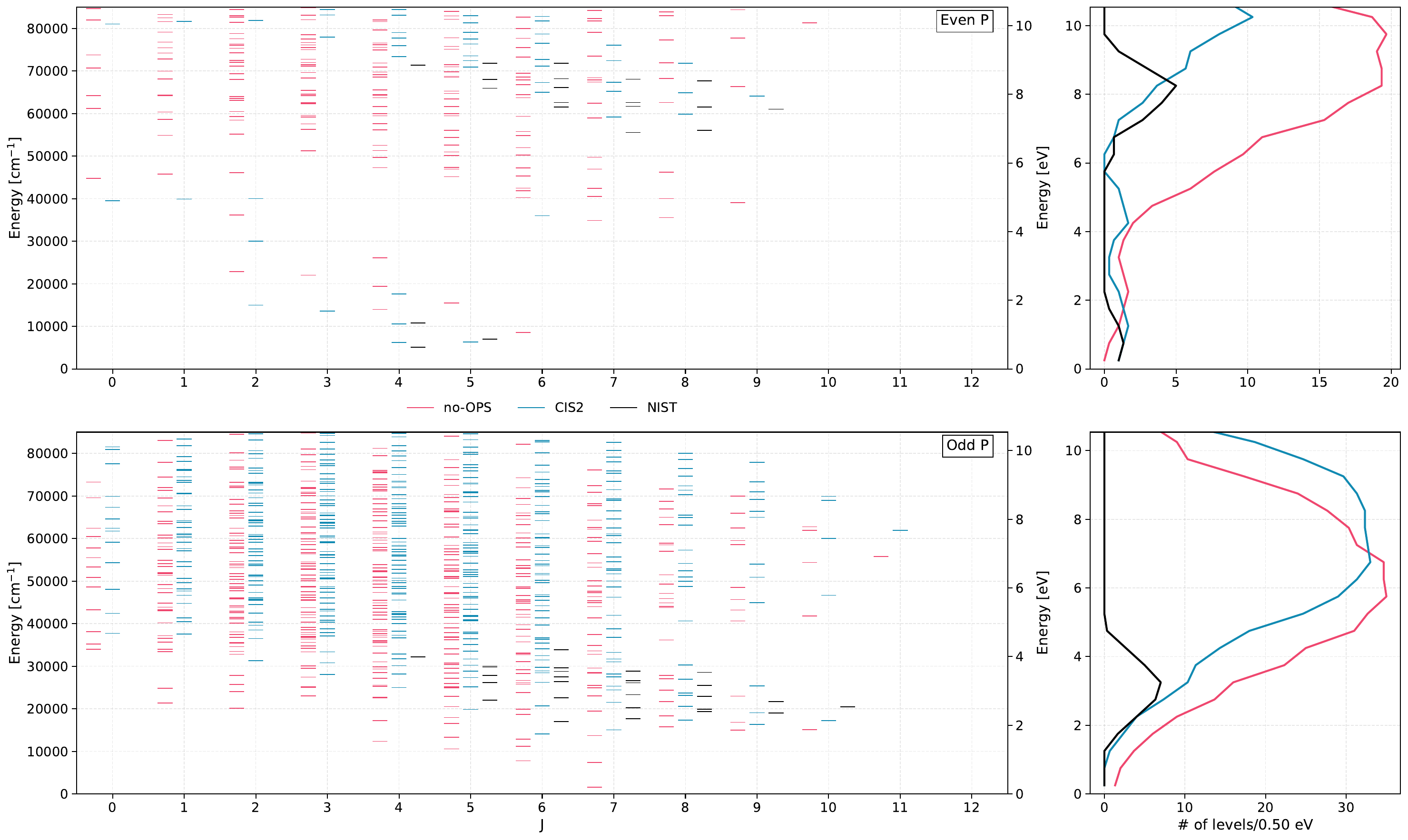}
\caption{Energy levels and level density of \ion{Er}{iii}. Pink lines show results for a calculation without optimisation of the FMC (using a set of configurations equivalent to OPS) while blue lines show the results for our largest calculation, including the optimisation. Black horizontal lines show the data from the NIST ASD.\cite{Kramida2021w}}
\label{fig:levels ErIII}
\end{figure*}

The energy levels calculated for \ion{Er}{ii} and \ion{Er}{iii} have been rigorously compared with the recommended values of the NIST database, as shown in \cref{fig:levels ErII} and  \cref{fig:levels ErIII}. Similarly to the results observed for \ion{Pr}{ii} and \ion{Pr}{iii}, the initial calculations using a configuration set equivalent to OPS reveal significant discrepancies from the NIST data. However, when FMC optimisation is applied, a notable enhancement in accuracy is achieved. Detailed results for the first few levels are shown in \cref{tab:energy_levels_erii,tab:energy_levels_eriii}.

In \cref{fig:levels ErII}, the optimised results for lower energy levels align closely with the NIST data, showing the effectiveness of the FMC optimisation. However, for higher energy levels, although there is improvement, discrepancies persist. This is particularly noticeable in the even parity levels, where the optimised results still show some deviations from the NIST data. In general, such discrepancies can create imbalances in the optimisation process, potentially leading to issues like overfitting, especially in odd levels. Including more configurations in the CI usually mitigates these discrepancies. Furthermore, the absence of the lowest levels for $J$ = 3, 5 and 7 for odd parity in the NIST database reduces the reliability of the comparisons for those levels.

In \cref{fig:levels ErIII}, the optimised results show that the level density is a much better match to the NIST data compared to the non-optimised case. This indicates that not only are individual levels shifted closer to the reference values but also the overall spectral density improves significantly with optimisation, even though only a few levels are found for even parity. However, it is important to note that this optimisation is highly dependent on the available data. Should missing levels be discovered in the range 0-60000 cm$^{-1}$ for even parity, the results may become inaccurate, necessitating the determination of a new FMC.

\begin{figure}
    \centering
    \includegraphics[width=\columnwidth]{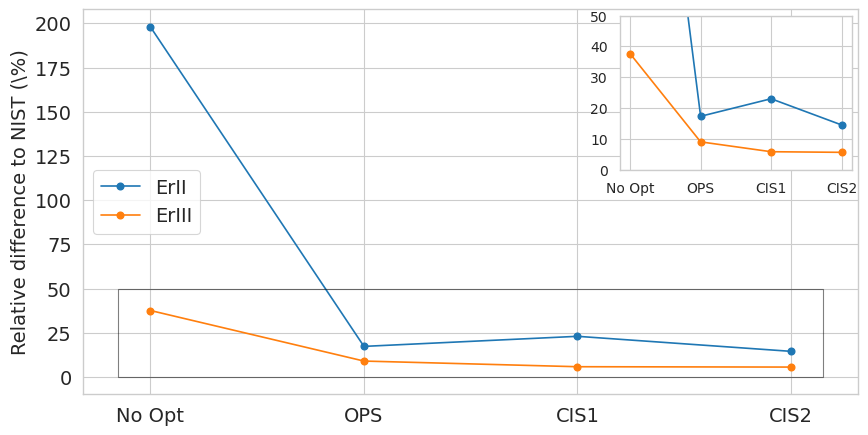}
    \caption{Average relative difference to the data available in the NIST ASD \cite{Kramida2021w} for all identified levels of \ion{Er}{ii} and \ion{Er}{iii} for all the different models computed with \FAC. ``No Opt.'' corresponds to an OPS configuration model that does not use the \texttt{FAC} potential optimisation, contrary to the other models.}
    \label{fig:levels er opti}
\end{figure}

The relative differences between our calculated energy levels and the NIST data for \ion{Er}{ii} and \ion{Er}{iii} across different computational models are depicted in  \cref{fig:levels er opti}. The ''No Opt." model, representing the OPS configuration without \FAC\, potential optimisation, shows higher relative differences compared to models that include this optimisation. The optimisation leads to a substantial reduction in relative differences, outperforming the use of just CI, as was also evident in the results for \ion{Pr}{ii} and \ion{Pr}{iii}. Moreover, examining the progression across different CI spaces (CIS1 and CIS2) shows that the optimisation maintains consistent improvements in accuracy, demonstrating the robustness of our approach.

\begin{table*}[ht]
\centering
\caption{Energy levels (in cm$^{-1}$) for the first 20 levels of \ion{Er}{ii}. The columns show: electronic configuration (Config.), level labels in LS and $jj$ coupling,, calculated energies without optimization ($E_{\text{no-OPS}}$), with optimization space only ($E_{\text{OPS}}$), and with increasing configuration interaction spaces ($E_{\text{CIS1}}$, $E_{\text{CIS2}}$). The experimental values from NIST ($E_{\text{NIST}}$) and relative differences between CIS2 and experimental values ($\Delta_{\text{CIS2}}$) are given in the last two columns.}
\begin{tabular}{lcccccccc}
\hline\hline
Config. & LSJ & $jj$ & $E_{\text{no-OPS}}$ & $E_{\text{OPS}}$ & $E_{\text{CIS1}}$ & $E_{\text{CIS2}}$ & $E_{\text{NIST}}$ & $\Delta_{\text{CIS2}}(\%)$ \\
\hline
$4f_{12} \, 6s$ & ${}^{4}H_{13/2}$ & $(6,1/2)_{13/2}$ & 16791.28 & 0.00 & 0.00 & 0.00 & 0.00 & --- \\
$4f_{12} \, 6s$ & ${}^{2}H_{11/2}$ & $(6,1/2)_{11/2}$ & 17556.49 & 515.66 & 661.75 & 442.86 & 440.43 & 0.65\% \\
$4f_{12} \, 6s$ & ${}^{2}G_{4}$ & $(4,1/2)_{9/2}$ & 23985.62 & 5652.03 & 5774.19 & 5238.72 & 5132.61 & 2.07\% \\
$4f_{12} \, 6s$ & ${}^{2}H_{3}$ & $(4,1/2)_{7/2}$ & 23057.16 & 5791.67 & 6053.75 & 5495.85 & 5403.69 & 1.71\% \\
$4f_{11} \, 6s^2$ & ${}^{4}I_{15/2}$ & $(15/2)_{15/2}$ & 6087.28 & 7190.58 & 9328.78 & 9347.81 & 6824.77 & 36.96\% \\
$4f_{12} \, 6s$ & ${}^{4}H_{11/2}$ & $(5,1/2)_{11/2}$ & 23944.58 & 7545.72 & 10790.22 & 7434.18 & 7149.63 & 3.98\% \\
$4f_{12} \, 6s$ & ${}^{2}H_{9/2}$ & $(5,1/2)_{9/2}$ & 22659.28 & 8208.46 & 8493.39 & 7590.97 & 7195.35 & 5.02\% \\
$4f_{11} \, 5d \, 6s$ & ${}^{4}F_{13/2}$ & $(15/2, 3/2, 1/2)_{13/2}$ & 9499.76 & 11788.31 & 11845.91 & 11160.24 & 10667.19 & 4.62\% \\
$4f_{12} \, 6s$ & ${}^{2}K_{7/2}$ & $(4, 1/2)_{7/2}$ & 27905.85 & 12532.39 & 12803.65 & 11486.81 & 10893.94 & 5.56\% \\
$4f_{12} \, 6s$ & ${}^{4}F_{9/2}$ & $(4, 1/2)_{9/2}$ & 28110.92 & 13867.35 & 13128.59 & 11615.88 & 11042.64 & 5.19\% \\
$4f_{11} \, 5d \, 6s$ & ${}^{4}H_{15/2}$ & $(15/2, 3/2, 1/2)_{15/2}$ & 10886.83 & 12942.23 & 14537.95 & 12035.95 & 11309.18 & 6.51\% \\
$4f_{11} \, 5d \, 6s$ & ${}^{4}F_{11/2}$ & $(15/2, 3/2, 1/2)_{11/2}$ & 12837.32 & 14064.20 & 14482.92 & 13194.92 & 12388.09 & 6.51\% \\
$4f_{12} \, 6s$ & ${}^{2}K_{7/2}$ & $(6, 1/2)_{7/2}$ & 30661.42 & 13848.06 & 13338.24 & 14148.52 & 12588.00 & 14.17\% \\
$4f_{12} \, 6s$ & ${}^{4}F_{5/2}$ & $(6,1/2)_{5/2}$ & 30660.60 & 14923.55 & 15829.49 & 14247.63 & 12600.09 & 13.09\% \\
$4f_{11} \, 5d \, 6s$ & ${}^{4}F_{19/2}$ & $(15/2,5/2)_{19/2}$ & 11191.85 & 15128.19 & 16687.78 & 13992.21 & 12815.07 & 9.21\% \\
$4f_{11} \, 5d^2$ & ${}^{2}F_{17/2}$ & $(15/2,2)_{17/2}$ & 12529.68 & 16320.09 & 13955.52 & 13921.83 & 13027.93 & 7.87\% \\
$4f_{11} \, 5d^2$ & ${}^{4}F_{13/2}$ & $(15/2,2)_{13/2}$ & 13089.82 & 17160.48 & 14199.61 & 15838.63 & 13060.72 & 21.28\% \\
$4f_{12} \, 6s$ & ${}^{6}H_{3/2}$ & $(2,1/2)_{3/2}$ & 31496.55 & 15922.44 & 14289.71 & 14762.72 & 13188.47 & 11.93\% \\
$4f_{11} \, 6^2$ & ${}^{4}I_{13/2}$ & $(13/2)_{13/2}$ & 13951.13 & 18536.90 & 15507.67 & 19659.70 & 13338.78 & 47.35\% \\
$4f_{12} \, 6s$ & ${}^{4}F_{1/2}$ & $(2,1/2)_{5/2}$ & 32244.40 & 18581.69 & 17380.42 & 14683.12 & 13558.33 & 8.30\% \\
\hline\hline
\end{tabular}
\label{tab:energy_levels_erii}
\end{table*}

\begin{table*}[ht]
\centering
\caption{Energy levels (in cm$^{-1}$) for the first 20 levels of \ion{Er}{iii}. The columns show: electronic configuration (Config.), level labels in LS and $jj$ coupling,, calculated energies without optimization ($E_{\text{no-OPS}}$), with optimization space only ($E_{\text{OPS}}$), and with increasing configuration interaction spaces ($E_{\text{CIS1}}$, $E_{\text{CIS2}}$). The experimental values from NIST ($E_{\text{NIST}}$) and relative differences between CIS2 and experimental values ($\Delta_{\text{CIS2}}$) are given in the last two columns.}\begin{tabular}{lcccccccc}
\hline\hline
Config. & LSJ & $jj$ & $E_{\text{no-OPS}}$ & $E_{\text{OPS}}$ & $E_{\text{CIS1}}$ & $E_{\text{CIS2}}$ & $E_{\text{NIST}}$ &$\Delta_{\text{CIS2}}(\%)$  \\
\hline
$4f_{12}$ & ${}^{3}H_{6}$ & $(6)_{6}$ & $8549.20$ & $0.00$ & $0.00$ & $0.00$ & $0.00$ & --- \\
$4f_{12}$ & ${}^{3}F_{4}$ & $(4)_{4}$ & $13986.81$ & $6379.26$ & $6426.10$ & $6187.68$ & $5081.79$ & 21.76\% \\
$4f_{12}$ & ${}^{3}H_{5}$ & $(5)_{5}$ & $15452.62$ & $7607.98$ & $7734.53$ & $6306.63$ & $6969.78$ & 9.51\% \\
$4f_{12}$ & ${}^{3}H_{4}$ & $(4)_{4}$ & $19390.33$ & $11699.33$ & $12083.23$ & $10558.93$ & $10785.48$ & 2.10\% \\
$4f_{11} \, 5d$ & ${}^{3}F_{6}$ & $(15/2,3/2)_{6}$ & $18680.31$ & $19945.03$ & $19118.33$ & $14071.95$ & $16976.09$ & 17.11\% \\
$4f_{11} \, 5d$ & ${}^{1}D_{7}$ & $(15/2,3/2)_{7}$ & $13700.74$ & $20431.81$ & $19052.84$ & $15053.80$ & $17647.76$ & 14.70\% \\
$4f_{11} \, 5d$ & ${}^{5}H_{9}$ & $(15/2,3/2)_{9}$ & $16825.73$ & $21728.27$ & $19570.55$ & $19060.32$ & $18976.74$ & 0.44\% \\
$4f_{11} \, 5d$ & ${}^{3}P_{8}$ & $(15/2,3/2)_{8}$ & $18330.78$ & $21319.04$ & $19606.51$ & $17325.56$ & $19315.90$ & 10.30\% \\
$4f_{11} \, 5d$ & ${}^{5}K_{8}$ & $(15/2,3/2)_{8}$ & $21683.07$ & $23223.31$ & $22518.24$ & $20539.81$ & $19918.17$ & 3.12\% \\
$4f_{11} \, 5d$ & ${}^{3}H_{7}$ & $(15/2,3/2)_{7}$ & $19491.41$ & $22174.29$ & $22707.52$ & $21519.74$ & $20226.20$ & 6.40\% \\
$4f_{11} \, 5d$ & ${}^{3}P_{10}$ & $(15/2,5/2)_{10}$ & $15098.56$ & $17639.37$ & $18084.44$ & $17200.91$ & $20470.13$ & 15.97\% \\
$4f_{11} \, 5d$ & ${}^{3}H_{9}$ & $(15/2,5/2)_{9}$ & $22969.92$ & $26168.92$ & $26496.96$ & $25397.46$ & $21688.17$ & 17.10\% \\
$4f_{11} \, 5d$ & ${}^{3}K_{5}$ & $(15/2,5/2)_{5}$ & $22888.83$ & $21061.99$ & $24145.93$ & $19851.71$ & $22016.77$ & 9.83\% \\
$4f_{11} \, 5d$ & ${}^{5}I_{6}$ & $(15/2,5/2)_{6}$ & $23823.06$ & $21723.57$ & $23138.71$ & $20662.19$ & $22606.07$ & 8.60\% \\
$4f_{11} \, 6s$ & ${}^{3}I_{8}$ & $(15/2,5/2)_{8}$ & $24388.94$ & $24292.88$ & $23495.02$ & $23135.29$ & $22951.42$ & 0.80\% \\
$4f_{11} \, 6s$ & ${}^{5}K_{7}$ & $(15/2,5/2)_{7}$ & $23044.05$ & $20474.16$ & $20383.00$ & $24433.67$ & $23302.78$ & 4.85\% \\
$4f_{11} \, 5d$ & ${}^{3}I_{8}$ & $(13/2,3/2)_{8}$ & $27084.63$ & $23071.16$ & $24660.08$ & $26950.80$ & $25482.12$ & 5.76\% \\
$4f_{11} \, 5d$ & ${}^{5}K_{7}$ & $(13/2,1/2)_{7}$ & $25501.29$ & $29257.19$ & $24928.37$ & $25305.76$ & $26102.80$ & 3.05\% \\
$4f_{11} \, 5d$ & ${}^{5}K_{5}$ & $(15/2,3/2)_{5}$ & $25940.92$ & $26752.22$ & $27523.70$ & $25187.40$ & $26192.66$ & 3.84\% \\
$4f_{11} \, 5d$ & ${}^{5}I_{6}$ & $(13/2,1/2)_{6}$ & $26115.01$ & $25562.72$ & $23877.65$ & $26226.36$ & $26411.80$ & 0.70\% \\
\hline\hline
\end{tabular}
\label{tab:energy_levels_eriii}
\end{table*}

\subsubsection{Transitions}
\begin{figure}
\centering
\includegraphics[width=\linewidth]{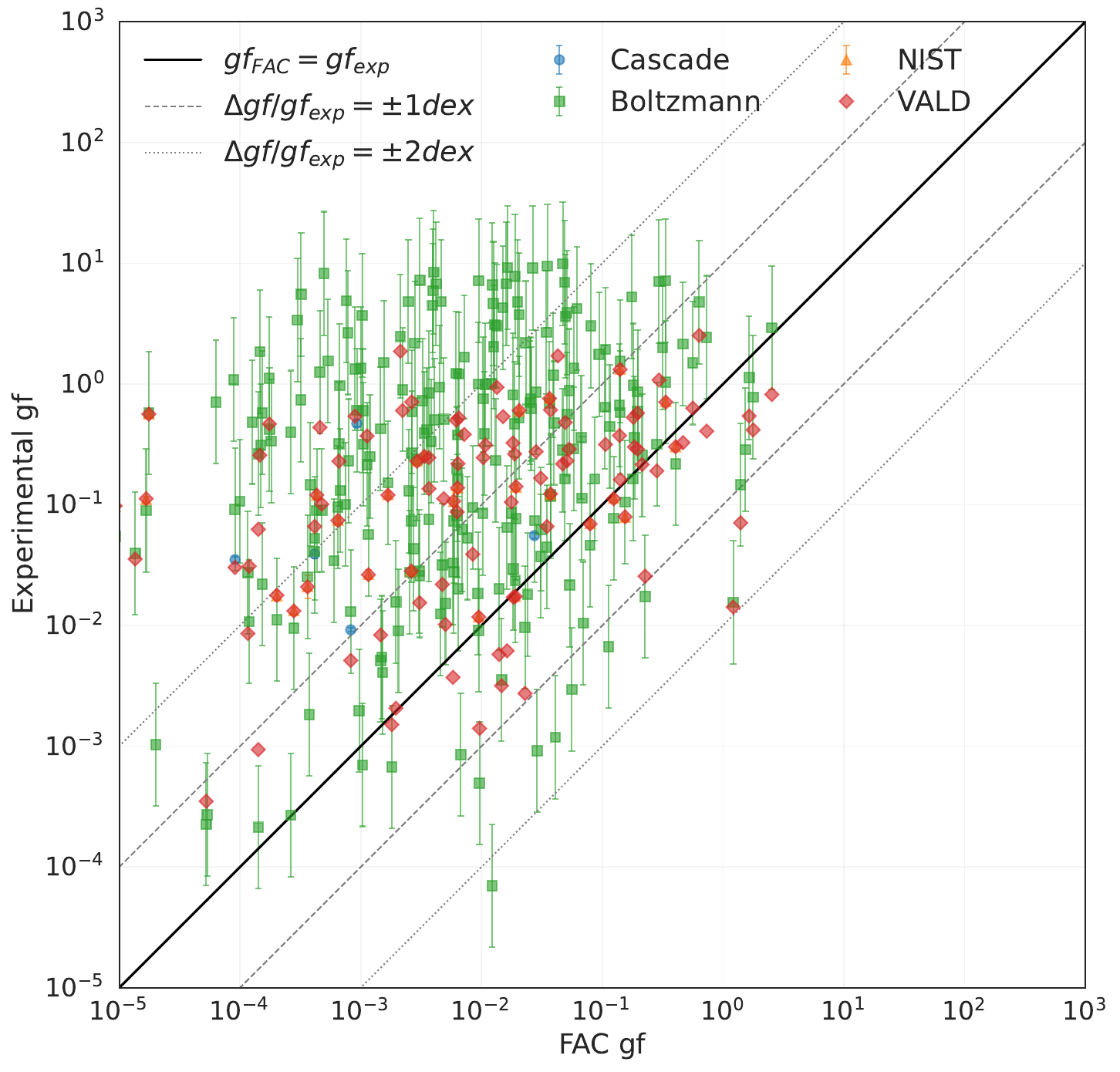}
\caption{Comparison of calculated $\log(gf)$ values from {CIS2} model with experimental data for \ion{Er}{ii}. The y-axis shows the experimental $\log(gf)$ values, while the x-axis represents the calculated values. Different colours indicate the source of the experimental data: cascade (blue circles) and Boltzmann plot method (orange triangles) available from Ferrara et al. (2024) \cite{Ferrara2023q}; VALD (red squares) and the NIST database (orange triangles).}
\label{fig:ergf}
\end{figure}
\begin{figure}
\centering
\includegraphics[width=\linewidth]{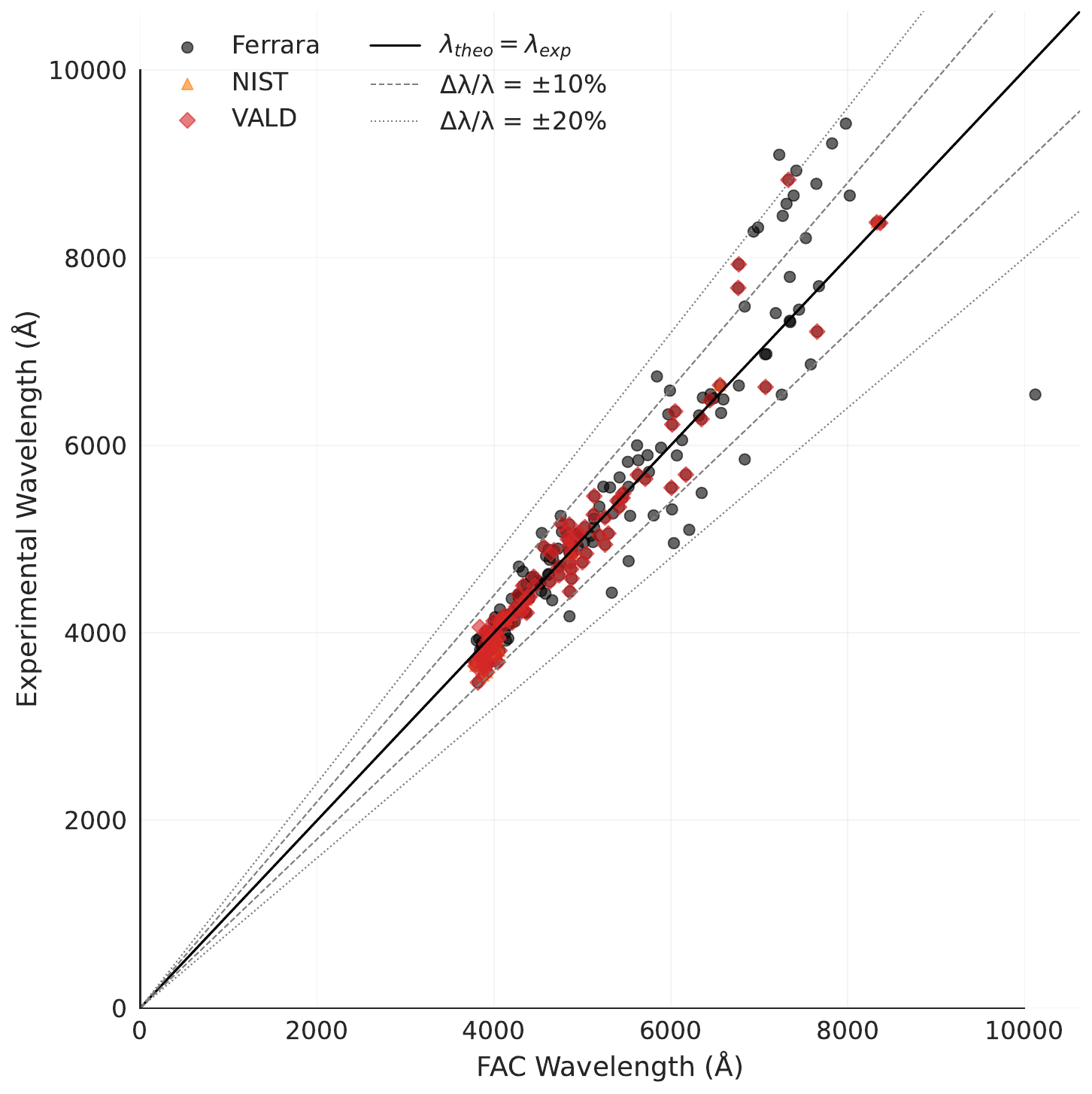}
\caption{Comparison of calculated wavelengths from ${CIS2}$ model with experimental data for \ion{Er}{ii}. The y-axis shows the experimental wavelengths in \AA ngstroms, while the x-axis represents the calculated wavelengths. Grey dots represent comparison to available experimental data from Ferrara et al. (2024) \cite{Ferrara2023q}, while orange triangles and red squares compare our data with data available in VALD and NIST ASD, respectively.  The dashed lines indicate a 10\% difference from the experimental data, while the dotted lines show a 20\% difference.}
\label{fig:erwv}
\end{figure}

To gauge the reliability of our results for \ion{Er}{ii}, we again utilized the extensive experimental data from Ferrara et. al. 2024 and available data from VALD and NIST ASD.  \cref{fig:ergf} illustrates the comparison between $\log(gf)$ values calculated using our {CIS2} model and the experimental data for \ion{Er}{ii}. As with the praseodymium data, the experimental values were obtained through both Boltzmann plots and using a cascade method. 

As evident from \cref{fig:ergf}, there is significant scatter in the $\log(gf)$ values throughout the range of transition strengths. The agreement between our calculations and the experimental data does not show a clear trend with transition strength, unlike what was observed for \ion{Pr}{ii}. Quantitatively, we observe a RMSD of approximately 1.4 dex between our calculated $\log(gf)$ values and the experimental data. This level of disagreement is comparable to what we found for \ion{Pr}{ii}.

In contrast,  \cref{fig:erwv} demonstrates good agreement between our calculated wavelengths and the experimental values for \ion{Er}{ii}. Most of the calculated wavelengths fall well within 10\% of the experimental data, as indicated by the points that lie between the inner dotted lines. Virtually all calculated wavelengths are within 20\% of the experimental values, falling between the dashed lines. Quantitatively, we find that approximately 85\% of our calculated wavelengths deviate by less than 10\% from the experimental values, and over 95\% are within 20\%.
The good agreement in wavelength calculations for both \ion{Pr}{ii} and \ion{Er}{ii} ions represents a significant improvement over standard calculations without optimization and approaches the level of accuracy typically achieved with more computationally intensive multiconfiguration Dirac-Fock (MCDF) methods like those implemented in the \texttt{GRASP} code \cite{Radziute2020n, Radziute2021s}. This level of accuracy is particularly relevant for opacity calculations in kilonova modelling, while requiring significantly less computational resources than full MCDF approaches.

The stark contrast in accuracy between the radiative transition probabilities and the wavelengths observed for \ion{Er}{ii} mirrors our findings for \ion{Pr}{ii}. This discrepancy can be attributed to the nature of our optimisation procedure, which is primarily designed to minimise differences in energy levels when compared to experimental data. While this approach typically leads to more accurate wavefunctions, these results clearly demonstrate its limitations. The high accuracy in wavelength predictions, which are primarily determined by energy level differences, is a direct consequence of this optimisation strategy. However, persistent discrepancies in oscillator strengths suggest that our current approach may not fully capture the intricacies of the wavefunctions required for accurate transition probability calculations. This indicates that including a larger basis set in our calculations may be necessary to improve the accuracy of oscillator strengths, as these are more sensitive to the fine details of the electronic wavefunctions.

\section{Systematic calculations}

With the aid of the optimisation technique described earlier, comprehensive calculations for multiple relevant r-process ions were conducted. According to kilonova models generated by radiative transfer codes for times \( t_{\text{exp}} \gtrsim 1 \) days, we focused on singly and doubly ionised ions due to their significant roles in the line-forming regions \cite{Shingles2023s}. Although higher ionisation stages are relevant at earlier stages and much higher temperatures, the impact of neutral species has been a recent topic of debate within the community. Because of their higher computational expense, calculations for relevant neutral ions are still under development. A notable use case of this optimisation technique was highlighted in a previous study, which focused on singly and doubly ionised Nd and U. This optimisation was performed manually using a uniform search grid of 100 values. The chosen FMC was the one that best reproduced the experimental energy of the lowest levels for each \( J \) and the parity value \cite{Flors2023a}.

\begin{table*}
\centering
\caption{Summary of optimization characteristics for singly ionized lanthanide ions. For each ion we list: the number of experimental levels used as reference data ($N_{\text{ref}}$), number of iterations required for convergence ($N_{\text{iter}}$), root mean square deviation (RMSD) between calculated and experimental energy levels before and after optimization. The reference data is taken from the NIST ASD. The improvement factor indicates the ratio of initial to final RMSD.}
\begin{tabular}{lcccccl}
\hline
Ion   & $N_{\text{ref}}$ & $N_{\text{iter}}$ & RMS$_{\text{initial}}$ (cm$^{-1}$) & RMS$_{\text{final}}$ (cm$^{-1}$) & Improvement factor \\
\hline
La II & 42  & 52  & 1418 &  673 & 2.11 \\
Ce II & 86  & 47  & 2384 & 1122 & 2.13 \\
Pr II & 97  & 54  & 3005 & 1601 & 1.88 \\
Nd II & 156 & 60  & 1552 & 1159 & 1.34 \\
Pm II & 38  & 48  & 5447 & 1276 & 4.27 \\
Sm II & 122 & 51  & 1304 & 1024 & 1.27 \\
Eu II & 89  & 50  & 4514 & 1422 & 3.17 \\
Gd II & 239 & 117 & 2598 & 1770 & 1.47 \\
Tb II & 72  & 103 & 6602 & 1200 & 5.50 \\
Dy II & 245 & 95  & 4141 &  706 & 5.87 \\
Ho II & 43  & 112 & 6219 & 1341 & 4.64 \\
Er II & 156 & 106 & 2092 & 1643 & 1.27 \\
Tm II & 168 & 124 & 2902 &  844 & 3.44 \\
Yb II & 142 & 67  & 3733 & 1767 & 2.11 \\
\hline
\end{tabular}
\label{tab:opt_performance_single}
\end{table*}

\begin{table*}
\centering
\caption{Summary of optimization characteristics for doubly ionized lanthanide ions. For each ion we list: the number of experimental levels used as reference data ($N_{\text{ref}}$), number of iterations required for convergence ($N_{\text{iter}}$), root mean square deviation (RMSD) between calculated and experimental energy levels before and after optimization. The reference data is taken from the NIST ASD, except for \ion{Dy}{III}, where the data is sourced from Spector et al (1997) \cite{Spector1997h}. No reliable experimental data for energy levels was found fro \ion{Pm}{III} besides for the ground state.  The improvement factor indicates the ratio of initial to final RMSD.}
\begin{tabular}{lcccccl}
\hline
Ion   & $N_{\text{ref}}$ & $N_{\text{iter}}$ & RMSD$_{\text{initial}}$ (cm$^{-1}$) & RMSD$_{\text{final}}$ (cm$^{-1}$) & Improvement factor \\
\hline
La III & 41  & 19 & 3200  & 1485  & 2.15 \\
Ce III & 224 & 34 & 5591  & 1482  & 3.77 \\
Pr III & 389 & 38 & 2717  & 1707  & 1.59 \\
Nd III & 30  & 22 & 2233  & 1895  & 1.18 \\
Pm III & 1   & 32 & 1690  & 1667  & 1.01 \\
Sm III & 58  & 25 & 9119  & 2814  & 3.24 \\
Eu III & 118 & 71 & 6254  & 1259  & 4.96 \\
Gd III & 258 & 68 & 11556 & 1876  & 6.16 \\
Tb III & 111 & 89 & 4653  & 4575  & 1.02 \\
Dy III & 107 & 91 & 5828  & 2647  & 2.20 \\
Ho III & 48  & 33 & 12601 & 1881  & 6.70 \\
Er III & 53  & 26 & 2360  & 1640  & 1.44 \\
Tm III & 129 & 29 & 1974  & 1513  & 1.31 \\
Yb III & 55  & 24 & 1459  & 678   & 2.15 \\
\hline
\end{tabular}
\label{tab:opt_performance_double}
\end{table*}

The calculations are performed in two primary stages. First, the FMC, used to determine the optimal potential, is computed using a reference dataset that includes accurately determined low-lying levels. The data from the NIST database was the primary reference for building this dataset \cite{Kramida2021w}. An exception was made for \ion{Dy}{III}, for which data from the experimental work of Spector et al. was used as a reference \cite{Spector1997h}. \Cref{tab:opt_performance_single,tab:opt_performance_double} summarise the optimisation characteristics and performance for single- and doubly ionised lanthanide ions, respectively. The improvement factors demonstrate the consistent effectiveness of our optimisation approach across the series, with particularly notable improvements for ions with well-characterised electronic structures.

Secondly, the basis space is increased to include all major configurations up to the ionisation energy. For lanthanides, it was found that considering SD excitation from the ground configuration up to \{$8s$, $8p$, $7d$\} effectively covers the most relevant configurations that directly contribute to opacity. Additional configurations, based on excitations up to \( n = 9 \) and \( \ell = 4 \), were included in the CI basis to improve the convergence of the results.

\begin{figure*}[h]
\centering
\includegraphics[width=\textwidth]{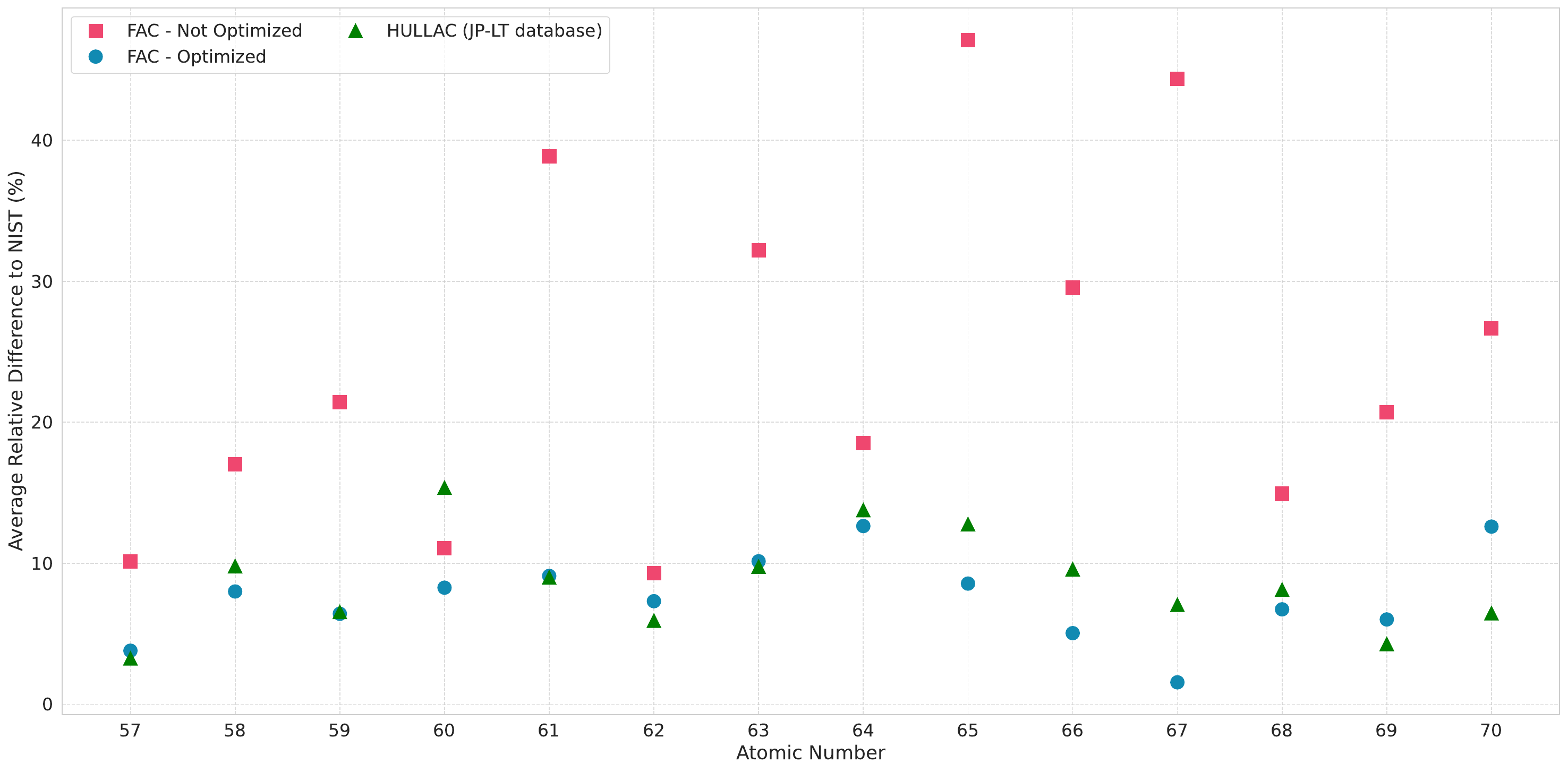} 
\caption{Average relative difference to the experimental data available in the NIST ASD \cite{Kramida2021w} for all singly ionized lanthanide ions computed with optimised FAC (in blue), compared with default \FAC\, (in red) and calculations from the Japan-Lithuania opacity database for kilonova (2021) obtained with \HULLAC\, (in green) \cite{KatoOthers}}
\label{fig:average_accuracy_single}
\end{figure*}

\begin{figure*}
\centering
\includegraphics[width=\textwidth]{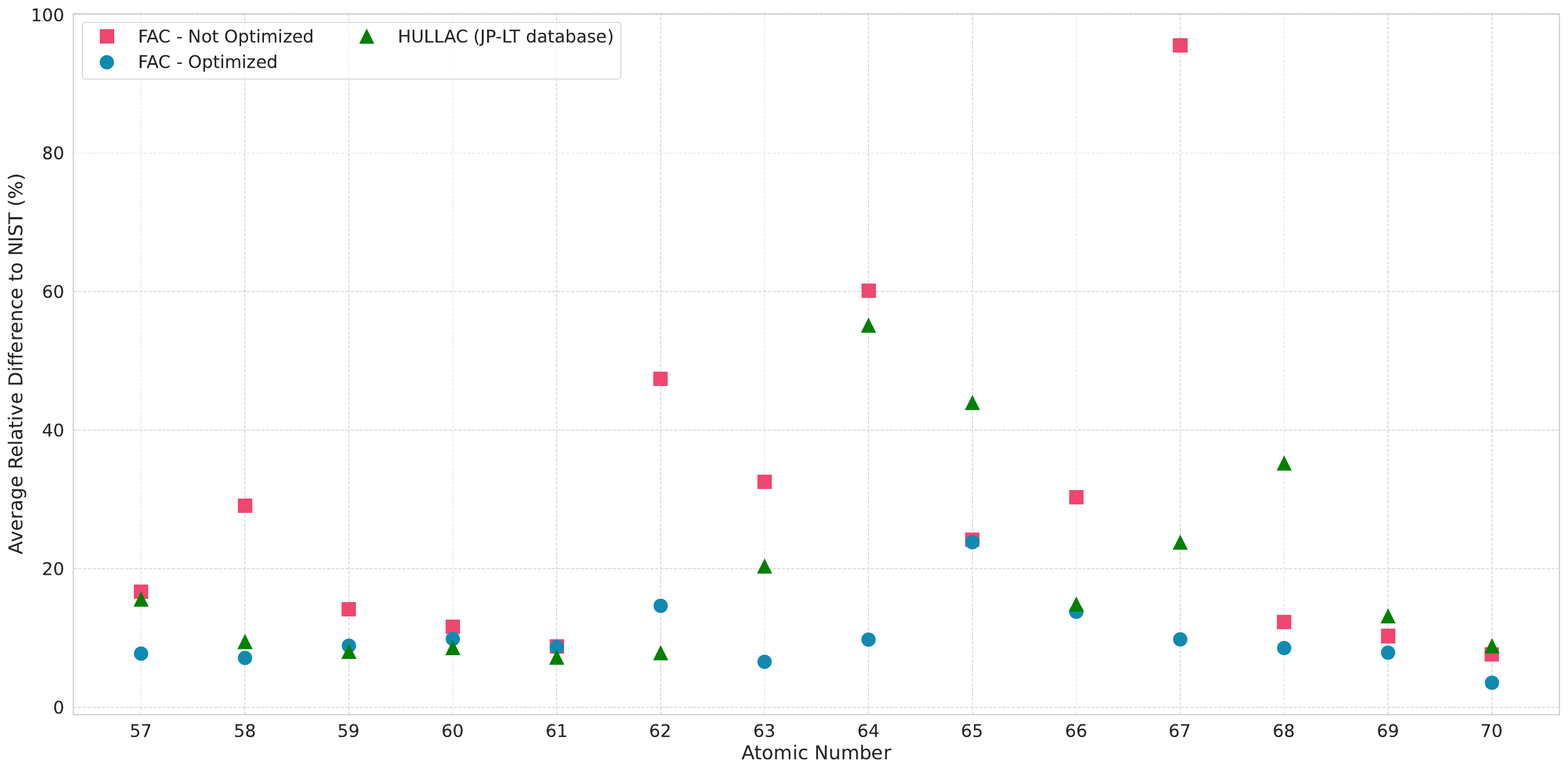}
\caption{Average relative difference to the experimental data available in the NIST ASD \cite{Kramida2021w} for all doubly ionized lanthanide ions computed with optimised \FAC\, (in blue), compared with default \FAC\, (in red) and calculations from the Japan-Lithuania opacity database for kilonova (2021) obtained with \HULLAC\, (in green) \cite{KatoOthers}}
\label{fig:average_accuracy_double}
\end{figure*}

\Cref{fig:average_accuracy_single,fig:average_accuracy_double} provide a summary of the average precision for single- and doubly ionised lanthanide calculations performed during this work. A difference of around 10\%, and in some cases lower, compared to the NIST data was found for both ionisation stages. These results align with, and in some cases improve upon, the systematic calculations performed with the \HULLAC\, code and available in the Japan-Lithuania opacity database for kilonova (2021). Additionally, unlike individual-level calibration, the optimisation of the central potential impacts the entire energy spectrum and, consequently, the level density. Although some differences have been observed compared to \textit{ab initio} calculations using the Hartree-Fock relativistic code (\texttt{HFR}), based on \cite{Cowan1981n} , the direct effect on opacity under local thermodynamical equilibrium conditions is expected to be minor, with more significant impacts at energies close to the ionisation energy.

Following the same calculation procedure used for the lanthanides, structure calculations were also performed for all singly- and doubly ionised actinides. In this case, because of the scarcity of actinide data in the NIST database, the reference energy levels for the optimisation of the FMC were sourced from the Selected Constants Energy Levels and Atomic Spectra of Actinides (SCASA) \cite{SCASA}, available as an online database. Whenever the number of experimental levels was insufficient for reliable optimisation, a fictitious configuration based on the FMC of the homologous lanthanide ion was constructed. The results and their consequent impact on opacities are currently being analysed, and publications are expected in the near future.

\section{Conclusions}
This study presents a novel approach to optimise atomic structure calculations using the \FAC\, for complex multielectron systems, with a particular focus on lanthanide and actinide ions relevant to $r$-process nucleosynthesis and kilonova modelling. Our method, centred on the optimisation of the fictitious mean configuration (FMC), has demonstrated significant improvements in the accuracy of calculated energy levels and transition properties for a range of elements and ionisation states.

The FMC optimisation technique has substantially improved the accuracy of the calculated energy levels for \ion{Au}{II}, \ion{Pt}{II}, \ion{Pr}{II}, \ion{Pr}{III}, \ion{Er}{II}, and \ion{Er}{III}. We observed a notable reduction in the average relative difference from NIST data, typically 20-60\% in unoptimized calculations to less than 10\% with our optimised approach. This improvement is particularly noteworthy for low-lying energy levels, which are crucial for accurate spectroscopic modelling.

Our optimised method has shown exceptional performance in calculating transition wavelengths. For both Pr and Er ions around 85-90\% of our calculated wavelengths fall within 10\% of experimental measurements. This high level of accuracy is essential for reliable identification and analysis of spectral features in astrophysical observations, particularly in the context of kilonova spectra.
While we observed significant improvements in energy level calculations and wavelength predictions, the computation of transition probabilities ($\log(gf)$ values) remains challenging. Although our optimised calculations show improvement compared to unoptimized results, notable discrepancies persist. These discrepancies highlight the inherent difficulties in accurately representing electron correlations in complex ionic systems, especially for lanthanides and actinides with open $f$-shells.

The effectiveness of our optimisation procedure across the lanthanide series demonstrates its broad applicability to elements with varying $4f$ shell complexities. This versatility is crucial for comprehensive modelling of r-process nucleosynthesis, where a wide range of elements play important roles. Our systematic calculations for singly and doubly ionised lanthanides have achieved accuracies comparable to, and in some cases better than, those obtained with other widely used codes like HULLAC, as evidenced by comparisons with the Japan-Lithuania opacity database for kilonova.

The success of this optimisation technique in improving the accuracy of atomic structure calculations has important implications for astrophysical modelling, particularly in the context of kilonova spectra and r-process nucleosynthesis. By providing more accurate atomic data, our work contributes to enhancing the reliability of opacity calculations and spectral modelling for heavy elements in extreme astrophysical environments. This improved accuracy is critical for interpreting observational data from kilonova events and understanding the production and distribution of heavy elements in the universe.

However, persistent challenges in calculating accurate radiative transition probabilities underscore the need for further refinement of our approach. Future work should focus on expanding the basis set to potentially improve the accuracy of oscillator strengths and transition probabilities. Although computationally demanding, investigating the inclusion of core-core and core-valence correlations may further enhance the accuracy of energy level and transition calculations. 
While a perfect agreement for all atomic energy levels simultaneously is challenging due to the limitations of the local central potential approximation, our method provides a consistent improvement across a wide range of levels, and we also expect it produces more reliable wavefunctions supporting more accurate level identification. This enhanced identification capability provides a solid foundation for future work involving detailed calibration of theoretical energy levels against experimental data \cite{Flors2025}.

Extending the optimisation technique to neutral species and higher ionisation stages is another important avenue for future research. This extension would allow for coverage of a broader range of astrophysical conditions, from the early, hot phases of kilonova evolution to later, cooler stages where lower ionisation states become relevant. Furthermore, applying this method to a wider range of elements, particularly actinides, will support a more comprehensive modelling of $r$-process nucleosynthesis.

\section*{Acknowledgements}

RFS acknowledges the support from National funding by FCT (Portugal), through the individual research grant 2022.10009.BD. RFS, JMS, and JPM acknowledge the support from FCT (Portugal) through research center fundings UIDP/50007/2020 (LIP) and UID/04559/2020 (LIBPhys), through project funding 2022.06730.PTDC,''Atomic inputs for kilonovae modeling (ATOMIK)" and, partly through project funding 2023.14470.PEX ''Spectral Analysis and Radiative Data for Elemental Kilonovae Identification (SPARKLE) \cite{SPARKLE2025}". GMP and AF acknowledge support by the European Research Council (ERC) under the European Union's Horizon 2020 research and innovation programme (ERC Advanced Grant KILONOVA No.~885281), the Deutsche Forschungsgemeinschaft (DFG, German Research Foundation) - Project-ID 279384907 - SFB 1245, and MA 4248/3-1, and the State of Hesse within the Cluster Project ELEMENTS. 

This work has made use of the VALD database, operated at Uppsala University, the Institute of Astronomy RAS in Moscow, and the University of Vienna.

\bibliography{references}

\end{document}